\documentclass[apJ]{emulateapj}

\usepackage{textcomp}
\usepackage{epsfig, epstopdf}
\usepackage{longtable,lscape}
\usepackage{amssymb,amsmath}

\newcommand{\CIV}{C~{\sc iv}}

\newcommand{\OVI}{O~{\sc vi}}

\newcommand{\NV}{N~{\sc v}}

\newcommand{\FeXXV}{\hbox{Fe {\sc xxv}}}
\newcommand{\FeXXVI}{\hbox{Fe {\sc xxvi}}}
\newcommand{\SXV}{\hbox{S {\sc xv}}}

\newcommand{\simgt}{\lower 2pt \hbox{$\, \buildrel {\scriptstyle >}\over {\scriptstyle\sim}\,$}}
\newcommand{\simlt}{\lower 2pt \hbox{$\, \buildrel {\scriptstyle <}\over {\scriptstyle\sim}\,$}}

\newcommand{\hs}{HS~0810+2554}

\newcommand{\chandra}{{\emph{Chandra}}}

\slugcomment{Received 2013 August 26; Accepted 2014 Jan 16}
\shorttitle{Ultrafast Outflow of AGN HS~0810+2554}
\shortauthors{CHARTAS ET AL.}

\begin{document}

\def\sarc{$^{\prime\prime}\!\!.$}
\def\arcsec{$^{\prime\prime}$}
\def\beginrefer{\section*{References}%
\begin{quotation}\mbox{}\par}
\def\refer#1\par{{\setlength{\parindent}{-\leftmargin}\indent#1\par}}
\def\endrefer{\end{quotation}}

\title{Magnified Views of the Ultrafast Outflow of the  $\MakeLowercase{z}$ = 1.51 AGN HS~0810+2554 }

\author{G. Chartas\altaffilmark{1,2}, F. Hamann\altaffilmark{3}, M. Eracleous\altaffilmark{4}, T. Misawa\altaffilmark{5}, M. Cappi\altaffilmark{6}, M. Giustini\altaffilmark{7}, J.~C.~Charlton\altaffilmark{4}, and M.~Marvin\altaffilmark{1}}

\altaffiltext{1}{Department of Physics and Astronomy, College of Charleston, Charleston, SC, 29424, USA, chartasg@cofc.edu}
\altaffiltext{2}{Department of Physics and Astronomy, University of South Carolina, Columbia, SC, 29208}
\altaffiltext{3} {Department of Astronomy, University of Florida, 211 Bryant Space Science Center, Gainesville, FL 32611-2055}
\altaffiltext{4}{Department of Astronomy \& Astrophysics, Pennsylvania State University, University Park, PA 16802}
\altaffiltext{5}{School of General Education, Shinshu University, 3-1-1 Asahi, Matsumoto, Nagano 390-8621, Japan}
\altaffiltext{6}{INAF-Istituto di Astrofisica Spaziale e Fisica cosmica di Bologna, via Gobetti 101, I-40129, Bologna, Italy}
\altaffiltext{7}{ESAC/ESA, PO Box 78, E-28691 Villanueva de la Canada, Madrid, Spain}

\begin{abstract}

\noindent We present results from an observation of the gravitationally lensed $z=1.51$ narrow absorption line AGN \hs\
performed with the  {\sl Chandra X-ray Observatory}. The factor of $\sim$100 lensing magnification of HS~0810+2554 
makes this source exceptionally bright.
Absorption lines are detected at rest-frame energies of  $\sim$ 7.7~keV and $\sim$11.0~keV at $\simgt$97\% significance. 
By interpreting these lines to arise from highly ionized iron the implied outflow velocities of the 
X-ray absorbing gas corresponding to these lines are 0.13 $c$ and 0.41$c$, respectively.
The presence of these relativistic outflows and the absence of any significant low-energy
X-ray absorption suggest that a shielding gas is not required for the generation of the relativistic X-ray absorbing winds in \hs.
UV spectroscopic observations with VLT/UVES indicate that the 
UV absorbing material is outflowing at $v_{\rm UV}$$\sim$0.065 $c$.
Our analysis indicates that the fraction of the total bolometric energy released by \hs\ into the IGM in the form of kinetic
energy is $\epsilon_{\rm k}$ = 1.0$_{-0.6}^{+0.8}$. 
An efficiency of greater than unity implies that magnetic driving is likely a significant 
contributor to the acceleration of this X-ray absorbing wind. 
We also estimate the mass-outflow rate of the strongest absorption component
to be  $\dot{M}_{\rm abs}$ = 1.1$_{-0.7}^{+0.9}$ M$_{\odot}$~yr$^{-1}$.
Assuming that the energetic outflow detected in the NAL AGN \hs\ 
is a common property of most AGN it would suggest that the X-ray absorbing wind may have a larger opening angle
than previously thought. This has important consequences for estimating the feedback contribution of X-ray absorbing 
winds to the surrounding IGM. 
\end{abstract}


\section{INTRODUCTION}

UV spectroscopic observations have revealed highly blueshifted intrinsic narrow 
absorption lines (NALs; FWHM $ $\simlt$ $500~km~s$^{-1}$) 
in several quasars implying outflow velocities of up to $\sim$
60,000~km s$^{-1}$ (e.g., Hamann, et al. 1997; Narayanan, et al. 2004; Simon et al. 2012).
NAL outflows are common in Type I AGN, occurring in $\sim$50\% of cases (Misawa
et al. 2007). They may be present in all AGN but only
detected in those cases where our line of sight intersects the
outflowing absorbing stream.  
Most of our current understanding of the physical and kinematic
structure of NALs stems from UV studies of the velocity profiles 
of absorption lines that appear blueward of resonance UV emission lines.
Little is presently known about the properties
of these absorption line systems in the X-ray band. 

Quasars with troughs $>$ 2000 km~s$^{-1}$ at 10\% depth below the continuum are traditionally referred to as BAL quasars (Weymann et al. 1991).
BAL quasars  are known to be heavily absorbed in X-rays
with intrinsic column densities in the range 10$^{23}$ -- 10$^{24}$~cm$^{-2}$
(Green \& Mathur 1996; Gallagher et al. 2002). 

Our X-ray survey of NAL AGN with fast outflows of UV absorbing material  
(Chartas et al., 2009a)  has revealed a remarkable X-ray bright object;
the $z$ = 1.51 gravitationally lensed AGN HS~0810+2554 was detected in a 5~ks $\it Chandra$ observation 
with a  2--10~keV observed  flux of {8.3$\times$10$^{-13}$~erg~s$^{-1}$~cm$^{-2}$}.
Its spectrum shows significant blueshifted  high energy absorption lines implying the presence 
of a massive and ultrafast X-ray absorbing wind with $v_{\rm X-ray}$  up to  0.41~$c$ (see \S3.1). 
UV spectroscopic observations with VLT/UVES indicate that the UV absorbing material of  
HS~0810+2554 is outflowing at $v_{\rm UV}$$\sim$19,400~km~s$^{-1}$ (see \S3.2; Culliton et al. in prep.).
The only other quasar detected to date with similar blueshifted X-ray absorption features
and a NAL outflow is the unlensed $z$=2.74 NAL quasar HS~1700+6414 (Lanzuisi et al. 2012)
which, however, has a much lower 2--10 keV  flux of {3--9}$\times$10$^{-14}$ erg~s$^{-1}$~cm$^{-2}$.
{\sl HST} observations of HS~0810+2554 with STIS revealed the quadruple lensed nature of this object (Reimers et al. 2002).
Asseff et al. (2011) have modeled the HS~0810+2554 gravitational lens system 
and estimated a magnification factor of about 100.
Assuming this magnification, HS 0810+2554 is a border line Seyfert/Quasar 
with an unlensed 2--10~keV luminosity of 5.8$\times$10$^{43}$ erg~s$^{-1}$.
Observations of  the $z = 1.51$ AGN \hs\ therefore provide us with the rare opportunity to spectroscopically study in detail 
the relativistic outflow of a low-luminosity NAL AGN  near the peak of the galaxy merger number density and cosmic AGN activity
and address many important science questions related to the nature of ultrafast ($v~\simgt$~10$^{4}$~km~s$^{-1}$) AGN outflows and their importance 
for feedback.

The X-ray absorption lines detected in \hs\ are similar to those found in other high-$z$ quasars (i.e., APM~08279+5255, Chartas et al. 2002; PG~1115+080, Chartas et al. 2003;  and HS~1700+6416, Lanzuisi et al. 2012).
The inferred hydrogen column densities of 
10$^{22}$--10$^{23}$~cm$^{-2}$
and near-relativistic velocities of these outflowing X-ray absorbers imply mass-outflow rates that are comparable to their estimated accretion 
rates. Systematic studies of the X-ray spectra of large samples of $z \simlt 0.1$ Seyferts (Tombesi, et al. 2010; Gofford et al. 2013) have revealed 
ultrafast outflows in $\simgt$ 40 \% of these samples.

The mechanism responsible for the acceleration of the X-ray absorbing material to near-relativistic velocities is 
currently debated.  The two main proposed mechanisms are radiation and magnetic driving
(e.g.,  Murray et al. 1995a, 1995b; Proga et al. 2000, 2004; Konigl \& Kartje 1994;  
Everett 2005, 2007; Fukumura et al. 2010, 2013).
In the radiation driving mechanism, radiation from the accretion disk and possibly corona accelerate
accretion disk material to near-relativistic speeds through Compton scattering, bound-free and bound-bound transitions.
In the magnetic driving mechanism, acceleration of the X-ray absorbers is due
to magneto-centrifugal forces generated by the rotation of the accretion disk and/or magnetic pressure forces. 
The wind acceleration may be produced by both magnetic and radiation driving mechanisms 
with their contributions varying as a function of distance from the black hole.
A crucial ingredient of the radiation driving mechanism is the postulated shielding gas (i.e., Murray et al. 1995)  
that prevents the outflowing gas from being over-ionized by radiation from the central source. 
The strength of the radiative driving force depends critically on both the column density of the shield  and its ionization level
(Chelouche \& Netzer 2003; Chartas et al. 2009b, Saez \& Chartas 2011).
The large intrinsic hydrogen column densities inferred in X-ray observations of BAL quasars have been proposed to represent the shielding gas (Gallagher et al. 2002, 2006) that is critical in the acceleration of the UV absorbing and X-ray absorbing gas.
However, the low-to-medium S/N of available X-ray spectra of BAL quasars has made it difficult to place 
useful constraints on the ionization properties of the X-ray absorbing gas in BAL quasars.

With a magnification factor of $\sim$100, \hs\  is among the X-ray brightest distant AGN for which, as shown in the analysis below, we have detected a relativistic 
outflow of X-ray absorbing material. 
In \S 2 we present the spectral and spatial analysis of  \hs\ and a nearby galaxy group that contributes to the lensing, in
 \S 3 we discuss the properties of the outflow and provide a test for the presence of a shielding gas,
 and in \S4 we present a summary of our conclusions.
Throughout this paper we adopt a flat $\Lambda$ cosmology with 
$H_{0}$ = 67~km~s$^{-1}$~Mpc$^{-1}$,  $\Omega_{\rm \Lambda}$ = 0.69, and  $\Omega_{\rm M}$ = 0.31, based on the Planck 2013 results 
(Planck Collaboration et al. 2013). The resulting luminosity distance to \hs\ is 11,405~Mpc.

\section{X-RAY OBSERVATION AND DATA ANALYSIS}
\hs\ was observed with the Advanced CCD Imaging Spectrometer (ACIS; Garmire et al. 2003) on board 
the {\sl Chandra X-ray Observatory} (hereafter {\sl Chandra}) on 
2002 January 30, with an effective exposure time 4894~sec. 
The pointing of the telescope placed 
\hs\ on the back-illuminated S3 chip of ACIS.
The {\sl Chandra} observation of \hs\ was analysed using the CIAO 4.5 software with 
CALDB version 4.5.5.1 provided by the {\sl Chandra X-ray Center} (CXC).
We used standard CXC threads to screen the data for status, grade, and time intervals of acceptable aspect solution and background levels. 
To improve the spatial resolution we employed the sub-pixel resolution technique developed by Li et al. (2004) and incorporated via the
Energy-Dependent Subpixel Event Repositioning (EDSER) algorithm into the tool acis\_process\_events of CIAO 4.5.

The \chandra\ spectrum of \hs\ was fit with a variety of models
employing \verb+XSPEC+ version 12 (Arnaud 1996).
For all spectral models of \hs\ we 
included Galactic absorption due to neutral gas (Dickey \& Lockman 1990) with a column density of 
 $N_{\rm H}$= 3.94 $\times$ 10$^{20}$~cm$^{-2}$.

\subsection{Spatial and Spectral Analysis of \hs}

We perform a spatial analysis of the {\sl Chandra} observation of \hs\ by fitting  the 
lensed images (see Figure 1) simultaneously using the relative astrometry derived from {\sl HST} images 
(Morgan et al. 2006) and \verb+MARX+ models (Wise et al. 1997) of the PSF.
We fit the data using 0\sarc0246 bins (compared to the 0\sarc491 ACIS pixel scale)
by minimizing the $C$-statistic (Cash 1979) between the observed and model images.
The PSF fitting method provided the best fit value and errors for the number of counts in each image, $N_{\rm i,psf}$ and 
$\sigma_{\rm i,psf}$, respectively. We define as \(N_{\rm ABCD,psf}  = \sum_{i} N_{\rm i,psf} \)
to be the total counts of all four images (A, B, C, and D) from the PSF fitting method.

While the PSF models should correctly recover the relative fluxes, errors in the PSF
models may bias the total fluxes. We corrected for this by renormalizing the 
total counts to match the total counts found in a 5\arcsec\ radius region
centered on the lens with a background correction based on an annulus from 
7\sarc5  to 50\arcsec\ around the lens. Specifically, the counts from each image as determined from 
the PSF fitting method were renormalized to $N'_{\rm i,psf}$ = R$\times$$N_{\rm i,psf}$, where 
R = $N_{\rm ABCD,ap}$/$N_{\rm ABCD,PSF}$ and $N_{\rm ABCD,ap}$ is the total counts from the aperture method. 
Table 1 lists the observation date, the observational identification number, the exposure time, 
and the $N'_{\rm i,psf}$ counts (in the 0.2--10~keV band) of each image.

Figure 1 shows a binned image, a deconvolved image and the best-fit PSF model of the {\sl Chandra} observation of \hs.
For the deconvolution we applied the Richardson-Lucy algorithm (Richardson 1972; Lucy 1974)
and supplied a point spread function (PSF) created by the simulation tool \verb+MARX+ (Wise
et al. 1997). Images A and B are not resolved, while image C is resolved in the deconvolution. Image D is likely too faint to
be reconstructed in the deconvolution.
The inferred 0.2$-$10~keV flux ratios of \hs\ from the best-fit PSF model are 
(A/D)$_{X-ray}$ = (8.5 $\pm$ 2.2), (B/D)$_{X-ray}$ = (7.0 $\pm$ 2.0), and (C/D)$_{X-ray}$ = (4.2 $\pm$ 1.1). 
For comparison, the optical HST V$-$band flux ratios of \hs\ are 
(A/D)$_{V-band}$ = 11.5 $\pm$ 0.6, (B/D)$_{V-band}$ = 5.6 $\pm$ 0.8, and (C/D)$_{V-band}$ = 4.0 $\pm$ 0.6. 
The X-ray and optical flux ratios are consistent within errors albeit obtained during different epochs.
This consistency implies that no significant magnification due to possible microlensing produced by stars in the lens galaxy is present 
during the X-ray observation in addition to the magnification of $\sim$ 100  that is produced by the gravitational potential of the lens.

The X-ray counts in the soft (0.2--2~keV) and 
hard (2--10~keV) bands for the combined images of \hs\ are 545 $\pm$ 23 and 164 $\pm$ 13, respectively.
The X-ray counts were extracted from a circle of radius 5~arcsec
centered on the mean location of the images.
The backgrounds were determined by extracting events within an
annulus centered on the mean location of the images
with inner and outer radii of 10~arcsec and 50~arcsec, respectively. 
The extracted spectra were grouped to obtain a minimum of 10 counts in each energy bin, 
close to the minimum required number of counts per bin for $\chi^2$ to be statistically valid
(e.g., Cash 1979; Bevington \& Robinson 2003).
We chose this grouping to satisfy this requirement and to allow the maximum spectral resolution for the low S/N spectrum of \hs.
To test the validity of the use of $\chi^2$ statistics in our analysis for our selected grouping of the data we
also used the $C$-statistic (Cash 1979)\footnote{The spectra were binned to have at least one count per bin.}
that does not have this limitation in binning the data.
The results obtained from fitting the data with
$\chi^2$ and $C$-statistics are consistent within the estimated error bars.
This provides additional support to the validity of our spectral fitting results. In the remaining of the paper we use the results 
from fits to the spectrum of \hs\ that employ the $\chi^2$-statistic (unless mentioned otherwise).

We performed fits to the spectrum of \hs\ using events 
in the 0.35--10~keV energy range with a variety of models of increasing complexity.
The minimum rest-frame energy of events included in our analysis
is therefore $\sim$ 0.89 keV.
The fit residuals show significant absorption at observed-frame energies of 1--5~keV.
To illustrate the presence of these features we fit the spectra in the observed-frame 
 5--10~keV band with a power-law model
(modified by Galactic absorption) and extrapolated this model to the energy ranges not fit
(see Figure 2).  
We proceed in fitting the following models to the data guided by the shape and location of  identified 
absorption residuals located at observed-frame energies of 1--5~keV:  
1) power-law modified by neutral intrinsic absorption (APL),
2) power-law modified by ionized intrinsic absorption (IAPL), 
3) power-law modified by neutral intrinsic absorption and four absorption lines (APL + 4AL), and
4) power-law modified by neutral intrinsic absorption and two outflowing intrinsic ionized absorbers (APL + 2IA).

The results from fitting these models to the \chandra\ spectrum of \hs\ are presented in  
Table 2.  No significant intrinsic neutral absorption is detected with fits using the APL model and we find intrinsic 
N$_{\rm H}$ $<$ 1.2 $\times$ 10$^{21}$~cm$^{-2}$ at the 99\% confidence level.
In Figure 3 we show the 68\%, 90\% and 99\% $\chi^{2}$ confidence contours of the 
intrinsic hydrogen column density, N$_{\rm H}$ versus 
the photon index assuming model 1 of Table 2.

It is possible that an intrinsic absorber is ionized and was therefore difficult to detect in the
$\sim$ 5~ks  {\sl Chandra} spectrum. 
With the IAPL model we investigated the presence of a possible mildly ionized intrinsic absorber
that could affect the low-energy (\simlt 1~keV observed-frame) continuum X-ray emission. Ionized absorbers 
have been postulated to provide shielding and prevent the outflowing gas from
being over-ionized (i.e., Murray et al. 1995).
Fits using the IAPL model do not provide useful constraints of the
column density in the case of ionized absorption at the redshift of the source (see Figure 3).
While the IAPL model provides an acceptable fit to the spectrum it does not provide any significant improvement over 
the APL model.
The inclusion of four Gaussian absorption lines near the absorption features  in the APL $+$ 4AL model 
(see Figure 2) resulted in a significant 
improvement of the fit compared to the fit using the APL model  
at the $ > $ 99.3\% confidence level (according to the $F$-test).
Specifically, the spectral fits indicate a change of $\chi^2$ per
change in degrees of freedom of	
$\Delta\chi^2/\Delta\nu$ = 20.3/12 and a change of the $C$-statistic per change in
degrees of freedom of $\Delta{C}/\Delta\nu$ = 23/12.
Protassov et al. (2002), argued that the $F$-test cannot be applied when
the null values of the additional parameters fall on the boundary of the
allowable parameter space. They proposed a Monte Carlo approach to determine
the distribution of the F statistic. We followed this approach and constructed the probability
density distribution of the $F$-statistic between spectral fits of models 1 and 3 of Table 2.
Our Monte Carlo simulations indicate that the probability of obtaining an $F$ value of
2.6 or greater is $\sim$ 5~$\times$~10$^{-3}$ similar to the result
found by using the analytic expression for the distribution of the $F$-statistic.
The best-fit observed-frame energies of the four Gaussian absorption lines are 
$E_{\rm abs1}$ = 1.6~keV, $E_{\rm abs2}$ = 2.0~keV,
$E_{\rm abs3}$  = 3.1~keV, and $E_{\rm abs4}$ = 4.5~keV (see Table 2).

We also investigated the improvement in the spectral fits of adding one Gaussian absorption
line at a time. The inclusion of one Gaussian absorption line near each of the four absorption features,
resulted in a slight improvement of the fit compared to the fit using the APL model.
Specifically, we find a change of $\chi^2$ per change in degrees of freedom of
$\Delta\chi^2/\Delta\nu$ = 5.19/2 for including a line near $E_{abs1}$,
$\Delta\chi^2/\Delta\nu$ = 4.20/2 for including a line near $E_{abs2}$,
$\Delta\chi^2/\Delta\nu$ = 4.15/2 for including a line near $E_{abs3}$, and
$\Delta\chi^2/\Delta\nu$ = 4.60/2 for including a line near $E_{abs4}$.
The respective $F$-values and the probabilities of obtaining $F$ values
equal to or larger than these values indicated by our Monte Carlo simulations
are $F$=3.1, $P$=0.1 for $E_{abs1}$,$F$=2.5, $P$=0.1 for $E_{abs2}$,
$F$=2.4, $P$=8 $\times$ 10$^{-2}$ for $E_{abs3}$,
and $F$=2.7, $P$=0.1 for $E_{abs4}$.

No significant intrinsic neutral absorption is detected with fits using the APL $+$ 4AL model and we find intrinsic 
N$_{\rm H}$ $<$ 0.93 $\times$ 10$^{21}$~cm$^{-2}$ at the 99\% confidence level, similar to the best-fit value of model 1.
To further test the robustness of our upper limit of the intrinsic neutral hydrogen column density, we took the
best-fit model 3 of Table 2  and forced the intrinsic column density to have a fixed value
of $N_{\rm H} = 5 \times 10^{21}~cm^{-2}$ for case 1 and  $N_{\rm H} = 1 \times 10^{22}~cm^{-2}$ for case 2, 
and redid the spectral fits for each case. In Figure 4 we show the best-fit models for each case
over-layed with the best-fit model where $N_{\rm H}$ was allowed to be a free parameter.
Figure 4 clearly shows that the hydrogen column densities assumed for cases 1 and 2 would lead to poor fits with significant
residuals in the 0.35--0.8 keV observed-frame energy band.
We conclude that our upper limit of the intrinsic hydrogen column density (assuming model 3 of Table 2)
is robust.
In Figure 5, we show the $\chi^2$ confidence contours of the rest-frame absorption line energies $E_{\rm abs}$ versus normalization of components abs1 ($E_{\rm abs1}$ = 3.94~keV), abs2 ($E_{\rm abs2}$ = 4.96~keV), 
abs3 ($E_{\rm abs3}$ = 7.74~keV), and abs4 ($E_{\rm abs3}$ = 11.0~keV) based on spectral fits that use the APL + 4AL model.
The 99\% confidence contours, of $E_{\rm abs1}$, $E_{\rm abs3}$, and $E_{\rm abs4}$ are erratic and not closed at the
99\% level and are therefore not displayed for clarity.

The number of background counts in the 5\arcsec\ source extraction circle are 3 counts in the 0.3-10~keV
energy range and 0.3 counts in the 1.5 - 5 keV range. We conclude that background subtraction is not the origin of
the four detected absorption lines in the 1.5 -- 5 keV energy range. 
There are several ACIS instrumental spectral features that fall near the observed-frame absorption line energies of 
$E_{\rm abs1}$ = 1.57~kV and $E_{\rm abs2}$ = 1.98~kV. Specifically, an Al K$\alpha$ absorption edge at  $\sim$1.56~keV from the ACIS
optical blocking filters and Ir M edge of the {\sl Chandra} X-ray mirrors at $\sim$ 2.085~keV.
We note, however, that the detected lines are too strong (see Table 2)
to be instrumental features and are not consistent with being absorption edges.
Specifically, we also modeled the four absorption features as absorption
edges using the \verb+zedge+ command in XSPEC. 
The spectral fits using absorption lines were better than fits that used
absorption edges, with a change of $\chi^2$ per
change in degrees of freedom of $\Delta\chi^2/\Delta\nu$ = 10.75/1, corresponding to a
$F$-statistic of 18.6.
Our Monte-Carlo simulations indicate that the probability of obtaining an $F$ value of
18.6 or greater is $\sim$ 1.3 $\times$ 10$^{-2}$ slightly larger than the value found
by the analytic expression for the distribution of the F-statistic.

The identification of the absorption lines at $E_{\rm abs1}$ and $E_{\rm abs 2}$ is ambiguous
and mostly speculative partly due to the detected rest-frame energies of the lines being below 6.7~keV,
the low S/N of the current spectrum and the relatively low-energy resolution available.
The identification of the absorption lines at $E_{\rm abs3}$  and $E_{\rm abs4}$ is less ambiguous because of their 
detected rest-frame energies being above 6.7~keV.
Specifically, the  high energy absorption lines that are likely associated with highly ionized iron are detected at rest-frame energies of 
$E_{\rm abs3}$ = 7.74~keV and $E_{\rm abs4}$ = 11.0~keV at $\simgt$ 97\% significance.
The best-fit results from the APL + 4AL model are used to identify the most significant absorption lines in \hs\
and provide a guide to fit an XSTAR photoionization model (APL + 2IA) containing two velocity systems to the spectrum
of \hs. Specifically, as our final refinement (model APL + 2IA) we model two of the high energy absorption features ($E_{\rm abs3}$ and $E_{\rm abs4}$) 
using the warm-absorber model \verb+XSTAR+ (Kallman et al. 1996;  Kallman \& Bautista 2001).
\verb+XSTAR+ calculates the physical conditions, and emission$\backslash$absorption spectra of photoionized gases.
In the current analysis we use an implementation of  the \verb+XSTAR+ model that can be used within
\verb+XSPEC+. 
Our fits with \verb+XSTAR+ indicate an outflowing X-ray absorbing medium with 
ionization parameters\footnote{Throughout this paper we
adopt the definition of the ionization parameter of Tarter et al. (1969) given by
$\xi=\frac{L_{\rm ion}}{n_H r^2}=\frac{4 \pi}{n_H}
\int_{1Rdy}^{1000Rdy}F_{\nu}d\nu$, where $n_H$ is the hydrogen number
density, and $r$ is the source-cloud separation.}
of components abs3 and abs4 of $\log\xi = 3.4^{+0.2}_{-0.2}$~erg~cm~s$^{-1}$
and $\log\xi = 3.1^{+0.2}_{-0.2}$~erg~cm~s$^{-1}$, respectively, assuming a $\Gamma$ = 2 ionizing continuum (model 4 of Table 2).
Of all the abundant elements, iron absorption lines would be the closest in energy to the observed features. 
In this sense, our interpretation that the absorption lines are associated with highly ionized Fe K absorption is the most 
conservative one possible (e.g., absorption lines from relativistic sulfur or oxygen would require much larger blueshifts).
A detailed justification of our interpretation that the high-energy absorption 
is due to lines arising from highly ionized \FeXXV\ is provided in 
related studies presented in Chartas et al. (2002, 2003, 2007) and Saez et al. (2007, 2011).
The two strongest iron lines for this highly ionized absorbing medium have rest (or laboratory)
energies of 6.70 keV (\FeXXV\ $1s^2-1s2p$) and 6.97 keV
(\FeXXVI\ $1s-2p$). In general the \FeXXV\ $1s^2-1s2p$
line will be stronger than the \FeXXVI\ $1s-2p$ line for a
medium with $2.75 \simlt \log\xi \simlt 4.0$ 
assuming an incident power-law spectrum with a photon index similar to the one inferred from fits using the APL + 2IA model
(see also Figure 3 of Saez et al. 2009).

For our spectral analysis we use the photo-ionization model \verb+XSTAR+ for the IAPL and
APL + 2IA models. We have selected to use the analytic \verb+XSTAR+ model \verb+warmabs+
instead of the \verb+XSTAR+ table models. According to the
\verb+XSTAR+ manual there are many advantages regarding accuracy and flexibility of using the analytic
\verb+XSTAR+  model instead of the \verb+XSTAR+ table model (Kallman \& Bautista \ 2001).
We emphasize, however,  that the photo-ionization model used in our analysis
does not consider possible velocity gradients in the outflowing absorber and therefore 
cannot provide realistic models of the X-ray absorption lines.

We attempt to mimic the velocity broadening of the lines by introducing in the
\verb+XSTAR+ \verb+warmabs+ model large turbulent velocities.
By introducing a turbulent velocity in  \verb+warmabs+,
the absorption line shapes are assumed by XSTAR to be Gaussian with
energy widths corresponding to the input turbulent velocity.
We performed several fits using  \verb+warmabs+
where we allowed the turbulence velocity to vary and found best-fit values of
$v_{\rm turb}$ $\sim$ 10,000~km~s$^{-1}$ for component abs3 and
$v_{\rm turb}$ $\sim$ 20,000~km~s$^{-1}$ for component abs4. 
Since there are several atomic transitions at slightly different energies
in addition to the strongest atomic transitions of Fe~xxv~1s2$-$1s2p and Fe~xxv~1s2$-$1s2p (e.g., see figure 3 of Chartas et al. 2009) 
we expect the best-fit values of the turbulent velocities to be slightly smaller than the velocity broadening inferred when
the features are modeled with only one absorption line as in model 3 of Table 3.
This is indeed what we find. Specifically, the widths of $\sigma_{\rm abs3}$$\sim$~300~eV and $\sigma_{\rm abs4}$$\sim$~970~eV (from model 3 using $C$-statistic 
of Table 2) correspond to velocity widths of 11,600~km/s and  26,000~km/s, respectively.
The velocity widths used in the  \verb+warmabs+ model are therefore consistent within errors
to the velocity widths implied by fits of single Gaussians to the absorption lines.
However, because of the low S/N of the {\sl Chandra} spectrum of \hs, the
turbulent velocities, are not well constrained.
For the error analysis of the remaining variables of model 4 of Table 3 we
therefore froze the turbulent velocities to $v_{\rm turb}$ = 10,000~km/s and
$v_{\rm turb}$ = 20,000~km~s$^{-1}$.
If we interpret the absorption
lines at rest-frame energies of 7.7$_{-0.2}^{+0.2}$~keV 
and 11.0$_{-0.8}^{+0.8}$~keV as being due to Fe, the most
conservative assignments (giving the lowest outflowing velocity) are to
highly ionized \FeXXV\ which requires $\log\xi \sim 3$ and/or highly ionized \FeXXVI\  which requires $\log\xi \simgt 4$, 
assuming an incident power-law spectrum with a photon index similar to the observed one.

In our spectral fits using the XSTAR photoionization code we are able to match the observed absorption lines
at E$_{\rm abs3}$ and E$_{\rm abs4}$ as \FeXXV\ outflowing at $0.13_{-0.09}^{+0.07}$~$c$ and $0.41_{-0.04}^{+0.05}$~$c$, 
respectively (see \S 3.1 for details).
If we allow the abundance of S to be a free parameter in the modeled outflow we are able to 
match the observed absorption line at E$_{\rm abs1}$ as arising from highly ionized \SXV\ ($1s^2-1s2p$)
(but at an abundance of $3_{-1}^{+2}$ the solar value)
outflowing at the same velocity of $0.41~c$ as the abs4 component.
We note that a detailed analysis of the physical conditions
of intrinsic narrow absorption line systems in three
quasars at $z = 2.6-3.0$ by Wu et al. (2010)
finds S to be overabundant relative to the Sun, consistent with our observations.

We note that the identification of the absorption lines abs1 and abs2 are not unambiguous
in our spectral analysis due to the low CCD energy resolution and the fact that there are many possible absorption lines
that fall in this energy range (see Figure 3 of Chartas et al. 2007). An X-ray spectrum of \hs\
with higher S/N and energy resolution will be required to	identify the absorption lines abs1 and abs2.
Our photoionization APL + 2IA model, that includes two outflowing components, does not 
indicate any clear absorption feature at E$_{\rm abs2}$.
One possibility is that the absorption line at E$_{\rm abs2}$ is produced by a third outflow or inflow component, however,
deeper follow-up X-ray observations are required to identify the origin of this spectral feature.
The spectra of the lensed images of HS 0810+2554 represent different epochs separated
by the time-delays between images. A deeper {\sl Chandra} observation can provide spatially
resolved and time-resolved spectra of the images (A+B), C, and D,
thus constraining the properties of the outflow in individual images.
AGN with relativistic outflows and with similar black hole masses show variability of the
X-ray absorption lines on timescales as short as 10~ks
(e.g., Chartas et al. 2007; Giustini et al. 2011; and Lanzuisi et al., 2012).
We thus expect variability of the high-energy absorption lines of HS~0810+2554 over similar timescales. A longer
observation will therefore provide insight into the variability of the properties of the outflow
in individual images. Our simulations based on the 5~ks {\sl Chandra} observation of \hs\ indicate that a  deeper
100~ks Chandra observation will provide a significant detection of the outflow at the $\simgt$ 99.99\% level and result
in improved constraints of the outflow properties in images (A+B) and C and a
detection of the blueshifted absorption lines in image D as well.
For example, a 100~ks {\sl Chandra} observation would result in the reduction of
the uncertainties of the detected energies of the absorption lines
by factors ranging between 2 and 4.

The column densities of the outflowing ionized components abs3 and abs4 (model 3 of Table 2) are 
N$_{\rm Habs3}$ = $1.7_{-0.6}^{+2.1}$ $\times$~10$^{23}$~cm$^{-2}$  and 
N$_{\rm Habs4}$ =  $1.6_{-0.9}^{+1.6}$ $\times$~10$^{23}$~cm$^{-2}$, respectively.
For comparison, the typical column densities of outflowing X-ray absorbing material in 
BALs range between 10$^{23}$~cm$^{-2}$ and 10$^{24}$~cm$^{-2}$ (e.g., Chartas et al., 2009b).
The column density of  outflowing X-ray absorbing material in the NAL quasar HS~1700+6416 
was found to be  $>$ 3 $\times$ 10$^{23}$~cm$^{-2}$ (Lanzuisi et al. 2012).

\subsection{Spatial and Spectral Analysis of the Group of Galaxies Near \hs}

In Figure~6 we show the {\it Chandra} image of the lensed system \hs\ and the surrounding field.
The image was binned with a bin size of 0.1 arcsec and adaptively smoothed with the CSMOOTH tool developed by 
Ebeling et al. (2000). CSMOOTH smooths a two-dimensional image with a circular Gaussian kernel of varying radius.
The binned data were smoothed using minimum and maximum smoothing scales of 0\sarc05 and 10\arcsec,
respectively. We detect extended X-ray emission centered at 
RA  = 8 13 27, Dec = $+$ 25 45 4.5 (J2000), approximately 58\arcsec\ west of \hs.
The diffuse X-ray emission detected 58\arcsec\ west of the lens is likely associated	with a group of	galaxies based	
on our analysis  of the spectrum of the extended X-ray emission presented below.
The NASA/IPAC Extragalactic Database (NED) lists nine $z \sim$ 0.08 galaxies within a separation of 5\arcmin\ of the lensing galaxy, 
one of which is separated by 9\arcsec\ from the center of the extended X-ray emission.
The overdensity of $z\sim0.08$ galaxies near the center of the extended X-ray emission supports our hypothesis
that the extended X-ray emission centered  58\arcsec\ west of the lens is associated with a group of galaxies
at $z = 0.08$. It is likely that the lensing galaxy is also a member of this group of galaxies.

We fit the emission from the foreground group
using a $\beta$ model for the group brightness profile (e.g., Jones \& Forman 1984) combined with a uniform background 
of 0.005 events per pixel.  Prior to performing the fit
we binned the image in 0\sarc5 pixels. 
The fits were performed with the CXC software package \verb+SHERPA+. 
We find that the group center is ${\Delta}{\alpha}$ = 58{\sarc}1   west
and ${\Delta}{\delta}$ = 1{\sarc}3  north of the center of \hs.
The intensity distribution is nearly round
with an ellipticity of ${\epsilon} = $ 0.15 $\pm$ 0.09.
The best-fit values for $\beta$ and the core radius of the group
are $\beta$ = 0.20 $\pm$ 0.01 and r$_{0}$ = 4{\sarc}9$_{-0.5}^{+0.8}$ (7.3~kpc),
respectively. 

We extracted the spectrum of this group 
from a 50 arcsec radius circle centered on the X-ray group center.
The extracted spectrum was of relatively low S/N 
containing a total of 98~counts
and was therefore fit using the $C$-statistic (Cash 1979)\footnote{The spectrum was binned to have at least one count per bin.}.
This spectrum was fit with a simple 
model consisting of emission from hot diffuse gas based 
on the XSPEC model {\it mekal} (Mewe, et al. 1985; Kaastra 1992; Liedahl, et al. 1995) modified by Galactic absorption. 
For display purposes only the spectrum was binned to have at least 5 counts per bin (Figure 7.). Our best-fit model is shown in Figure ~7. 
We obtain best-fit values (based on fits using the $C$-statistic) for the temperature and metal abundances
of  T$_{\rm e}$ = 1.06$_{-0.15}^{+0.15}$~keV and A = 0.5$^{+0.9}_{-0.3}$ solar,
respectively (both errors are at the 90\% confidence level).
The 0.2--10~keV and 2--10~keV luminosities of this galaxy group are 1.2 $\times$ 10$^{42}$~ergs~sec$^{-1}$
and 1.6 $\times$ 10$^{41}$~ergs~s$^{-1}$, respectively. The detected L$_{\rm X}$ and T$_{\rm e}$ for the galaxy group near \hs\ is 
consistent with the L$_{\rm X }$ $-$ T$_{\rm e}$ scaling relation for a  statistically  complete sample of galaxy groups (e.g., Eckmiller et al. 2011).

\subsection{Mass Distribution, Convergence and Shear of Galaxy Group Near \hs}

Assuming a $\beta$ model for the density profile of the hot gas
(e.g., Jones \& Forman 1984) the equation for the total mass of the galaxy group within a radius $r$,

\begin{equation}
{M_{grav}(<r)}  = { {{3 \beta k T}\over{\mu m_{p} G}} { {r} \over {[1 + ({{r_c}\over{r}})^2]}}} {\  ,}
\end{equation}

\noindent 
where $k$ is Boltzmann's constant, $m_{p}$ is the proton mass, and $\mu$ is the mean molecular weight of the galaxy group gas.

By incorporating the best-fit spatial and spectral parameters
from our present analysis we find the 
total galaxy group mass within radii of $r_{500}$
and 1~$h_{70}$$^{-1}$~Mpc to be 
M$_{grav}$ = 1.4$_{-0.3}^{+0.1}$ $\times$ 10$^{13}$ M$_{\odot}$
and M$_{grav}$ = 2.2$_{-0.3}^{+0.2}$ $\times$ 10$^{13}$ M$_{\odot}$, respectively, where
$r_{500}$ is the radius in which the mean over-density is 500,
and $r_{500}$ = 1.69(T$_{X}$/10~keV)$^{1/2}$~$h^{-1}_{70}$ Mpc $\sim$ 0.55 $h^{-1}_{70}$ Mpc
(Mohr, Mathiesen, \& Evrard, 1999).
In Figure 8 we show the total group mass within a radius $r$
as a function of radius. The shaded region indicates the allowed
values for the group mass including the uncertainties
obtained from the spatial and spectral fits to the group.

These mass estimates were used to evaluate the convergence
parameter ${\kappa}(x)$,

\begin{displaymath}
{\kappa(x) = {{\Sigma(x)}\over{\Sigma_{cr}}}} \,
\end{displaymath}
where ${\Sigma(x)}$ is the surface mass density of the galaxy group
as a function of the cylindrical radius $x$ (e.g., Chartas et al. 1998) and
$\Sigma_{cr}$ is the critical surface mass density 
(see, e.g., Schneider, Ehlers \& Falco 1992).
In Figure 9 we plot the convergence parameter $\kappa(x)$
as a function of distance from the group center.
The thick solid line corresponds to the best-fit spatial and spectral parameters.
The largest contributor to the uncertainty in our estimate 
of $\kappa(x)$ is the weak constraint on the temperature of the group.
To illustrate this we have plotted the uncertainty in
$\kappa(x)$ assuming  90\% (dashed lines) 
confidence intervals for the temperature. 
We also chose galaxy group limits ranging from 0.7$r_{500}$ and 1.4$r_{500}$.

At the location of the center of \hs\ with respect to the group center, 
we estimate the convergence due to the galaxy group to be ${\kappa_c}$ = 0.017$^{+0.011}_{-0.002}$, 
assuming the 90\% confidence range in group temperature.
The second order term of the Taylor series expansion 
of the potential of the galaxy group (Kochanek 1991, Bernstein \& Fischer 1999, Keeton et al. 2000)
represents the shear from the galaxy group that can be expressed as 
${\gamma} = {{\kappa}\over{({1 +{\beta_{rd}}^2})^{3/2}}}$ 
where, ${\beta_{rd}} = r_{c}/d_{c}$, $r_{c}$
is the galaxy group core radius and $d_{c}$ is the distance from the
group center to the center of \hs\ (see, Kochanek, 1991). 
Using the values for $r_{c}$, $d_{c}$, and
${\kappa_c}$ provided by our analysis of the {\sl Chandra} observation of
\hs\ we find that the group shear amplitude is $\gamma$ = 0.016
$\pm$ 0.003 (90\% confidence level).

\subsection{Lens Modeling of  \hs\ }

For estimating the energetics of the outflow of \hs\ it is important to constrain the unlensed luminosity
of the quasar and thus the magnification of each image. The galaxy group contributes to the lensing of the quasar
and our lens modeling of \hs\ provides estimates of the magnification and time-delays of the system
taking into account the perturbation effects of the galaxy group.

We used the gravitational lens adaptive-mesh fitting code glafic version 1.1.6 (Oguri 2010) to model
the gravitational lens system \hs.
The lens was modeled as a singular isothermal ellipsoid plus an external shear from the nearby galaxy group.
The ellipsoid's orientation and ellipticity were left as free parameters.

The external shear was constrained by our analysis of the \chandra\ observation of the galaxy group near \hs\
as presented in \S 2.3. We find that the magnifications of the images are $M_{\rm A}$=40, $M_{\rm B}$=38, $M_{\rm C}$=10, and $M_{\rm D}$=6,
consistent with Assef et al. (2011). 
The time-delays between images  are found to be $td_{\rm {BA}}$=86~s, $td_{\rm BC}$=6,220~s, 
$td_{\rm AC}$=6,307~s, and $td_{\rm DC}$=25,230~s.

\section{DISCUSSION}

\subsection{Mass-Outflow Rate and Efficiency of the Outflow}
Our analysis of the {\sl Chandra} observation of \hs\
places constraints on the velocity, column density and ionization parameter of the 
outflowing X-ray absorbing material. This allows us to estimate the mass-outflow rate and the efficiency of outflow from the 
expressions:

\begin{equation}
 \dot{M_{\rm i}} = 4{\pi}r_{\rm i}(r_{\rm i}/{\Delta}{r_{\rm i}})N_{\rm H,i}m_{p}v_{wind,i}f_{c,i}
\end{equation}

\begin{equation}
\epsilon_{\rm k,i} = {{1}\over{2}}{\dot{M}_{\rm i}{v^{2}_{\rm wind,i}}\over{L_{\rm Bol}}} 
\end{equation}

\noindent
where ${\Delta}{r_{\rm i}}$ is the thickness of the absorber at radius $r_{\rm i}$, 
$N_{\rm H,i}$ is the hydrogen column density,
$v_{\rm wind,i}$ is the outflow velocity of the X-ray absorber, $f_{\rm c,i}$
is the global covering factor of the absorber, $i$ indicates the absorbing component and $L_{\rm Bol}$ is the bolometric photon luminosity of the quasar.

The formalism that we adopt for estimating the mass-outflow rate assumes a spherically symmetric wind with a covering factor of
$f_{\rm c,i}$ (e.g., Lamers \& Cassinelli 1999).
We approximate $N_{\rm H} \sim n(r){\Delta}r$, where $n(r)$ is the number density
of the gas. For the special case of ${\Delta}r/r $ = 1 we obtain the equation used in Crenshaw \& Kraemer (2012).
A different formalism for estimating the mass-outflow rate proposed by Krongold et al. (2007)
applies to particular outflows with narrow biconical geometries as shown in their Figure 12.
We do not adopt this formalism because the assumptions made in that derivation likely do not apply to
relativistic outflows in NAL quasars.
We find the bolometric luminosity of \hs\ to be 
$L_{\rm Bol}$=2.4$\times$10$^{45}$(94/$\mu$)($\kappa_{2-10keV}$/30)~erg~s$^{-1}$
where $\kappa_{2-10keV}$ is the 2$-$10~keV bolometric correction factor and 
$\mu$ is the flux magnification.
We note that the bolometric luminosity of \hs\ is not well constrained because of the uncertainties in the lensing magnification and 
bolometric correction factor. An independent estimate of $L_{\rm Bol}$ is provided from the measured black hole mass of 
M$_{\rm BH}$ = 4.2 $\times$ 10$^{8}$ (94/{$\mu$})$^{0.5}$ M$_{\odot}$ and 
the dependence of the median Eddington ratio on the median M$_{\rm BH}$ for type-I AGN
found in XMM-COSMOS survey  (Lusso et al. 2012).  Based on the Lusso et al. results 
the mean Eddington ratio for an AGN with the black hole mass and redshift of  \hs\ is $\lambda_{\rm Edd}$ $\sim$ 0.1 
resulting in  $L_{\rm Bol}$=5.3~$\times$~10$^{45}$ (94/{$\mu$})$^{0.5}$ (0.1/$\lambda_{\rm Edd}$)~erg~s$^{-1}$.

We estimated the hydrogen column densities $N_{\rm H}$ of the X-ray absorption lines
from the fit of a photoionization model to the spectrum of \hs\ (see \S 2.1 and model 4 of Table 2).
From our spectral fits using model 4 we find 
N$_{\rm Habs3}$ = $1.7_{-0.6}^{+2.1}$$\times$~10$^{23}$~cm$^{-2}$  and N$_{\rm Habs4}$ =  1.6$_{-0.9}^{+1.6}$$\times$~10$^{23}$~cm$^{-2}$.
We note that in this model we assumed full coverage of the source by the outflowing absorbers.

The bulk outflow velocities of each outflow component were inferred from the best-fit values of the
redshifts of these components assuming model 4 of Table 2.
We find the outflow velocities of X-ray absorbers abs3 and abs4 to be 
$v_{\rm abs3}$ =  0.13$_{-0.09}^{+0.07}$~c and $v_{\rm abs4}$ =  0.41$_{-0.04}^{+0.05}$~c, respectively.

The global covering factor of the outflowing absorber is not constrained with the current observation.
A recent spectroscopic study along multiple lines of sight through the outflowing wind of the lensed quasar SDSS~J1029+2623
found that the column density of outflowing mini-BAL and NAL absorbers did not 
change significantly in a small angular separation of $\sim$ 20\arcsec\ (Misawa et al. 2013). 
In previous X-ray analyses of BAL quasars (e.g., Chartas et al. 2009b) we assumed a conservatively wide range for the 
covering factor of $f_{\rm c}$=0.1--0.3 
based on the observed fraction of BAL quasars (e.g., Hewett \& Foltz 2003). 
The detection of near-relativistic outflows in NAL quasars
HS~0810+2554 and  HS~1700+6414, however, would suggest that the X-ray absorbing winds in NAL quasars may 
have a larger opening angle that previously thought. Specifically, several studies of NAL quasar (e.g., Misawa et al. 2007; Simon et al. 2012;
Culliton et al. 2012) indicate that the true fraction of quasars with a quasar-driven outflow NAL is $\simgt$ 40\%. For our estimate 
of the mass-outflow rate we assumed a conservative wide range for the covering factor of $f_{\rm c}$=0.3--0.5.

For estimating the mass outflow rate and outflow efficiency, we assumed a fraction  $r/{\Delta}{r}$ ranging from 1 to 10 based on 
theoretical models of quasar outflows (e.g., Proga et al. 2000). Assuming that the maximum outflow 
velocity is produced by gas that has reached its terminal velocity one obtains the approximation
$R_{\rm launch}$ $\sim$ a few times $R_{s \rm}(c/v_{\rm wind} )^{2}$ , where $v_{wind}$ is the observed outflow velocity and $R_{\rm s}$ = $2GM/c^2$. 
Based on our estimated maximum outflow velocity ($v = 0.41c$), we expect 
$r$ to be similar to  $R_{\rm launch}$ and range between 3~$R_{\rm s}$ and 10~$R_{\rm s}$.
Our assumed range of $r$ is conservative, ranging from the innermost stable circular orbit (ISCO) radius to 10~$R_{\rm s}$.
Launching a wind at significantly smaller radii than the ISCO radius is unlikely
since general relativistic (GR) effects on an absorber launched within the ISCO radius would result in significant GR
redshifts of the absorption lines and launching radii greater than 10~$R_{\rm s}$ will result in even
larger values of the mass-outflow rate.

We used a Monte Carlo approach to estimate the errors of $\dot{M}_{\rm i}$ and $\epsilon_{\rm k}$.
The values of $v_{\rm wind}$ and $N_{\rm Habs}$ were assumed to have normal distributions within their error limits.
The values of $f_{\rm c}$, $r/{\Delta}{r}$, and $r$ were assumed to have uniform distributions within their error limits.
By multiplying these distributions and with the appropriate constants from equations 2 and 3 we obtainedÊ
the distributions of $\dot{M}_{\rm i}$ and $\epsilon_{\rm k}$.
We finally determined the means of the distributions of 
$\dot{M}_{\rm i}$ and $\epsilon_{\rm k}$ and estimated the 68\% confidence ranges.

In Table 3 we list the total hydrogen column densities $N_{\rm H}$ of the X-ray absorption lines,
the outflow velocity of each absorption component, the mass-outflow  rates and the efficiency of the outflows.
One intriguing result of our study is that the efficiency of the abs4 outflowing component is found to be 
$\epsilon_{\rm k}$ = 1.0$_{-0.6}^{+0.8}$(0.1/$\lambda_{\rm Edd}$).
An efficiency of greater than the covering factor $f_{\rm c}$ implies that radiation driving alone cannot explain the 
acceleration of this outflowing  highly-ionized absorber. It is likely that magnetic driving is a significant 
contributor to the acceleration of this X-ray wind (e.g., Kazanas et al. 2012;  Fukumura et al. 2013).
The mass outflow rate of the abs4 outflowing component is found to be 
$\dot{M}_{\rm abs4}$ = 1.1$_{-0.7}^{+0.9}$ M$_{\odot}$~yr$^{-1}$, which is comparable to the 
accretion rate of \hs\  which we estimate
to be 1.8 $\times$ 10$^{-3}$($L_{44}$/$\eta$)$M_{\odot}$  yr$^{-1}$ $\sim$ 1 $M_{\odot}$ yr$^{-1}$,
where we assumed a typical accretion efficiency of $\eta$ = 0.1.

\subsection{Testing the Shielding Hypothesis of AGN Outflows}
The radiation driving model of Murray et al. (1995) posits the presence of a shielding gas that is required to
drive the gas radiatively from small radii. The X-ray spectra of NAL AGN HS~0810+2554 (presented in this study) and HS~1700+6416
(presented in Lanzuisi et al. 2012) indicate that ultrafast outflows can be present in quasars without the 
presence of any significant absorbing shielding medium along the X-ray line-of-sight. 
Specifically, the ultrafast outflows detected in HS~1700+6416 and HS~0810+2554 are manifested as blueshifted absorption lines
in the X-ray spectra. These absorption lines have been interpreted as arising from highly ionized (log $\xi$  $\sim$ 3 erg~cm~s$^{-1}$) outflowing gas
with column densities of about N$_{\rm H}$ $\sim$ 10$^{23}$ cm$^{-2}$. 
This outflowing gas is too ionized and transparent to contribute to its shielding. 
Our spectral analysis of these two objects 
indicated the absence of any significant absorber that could act to shield the outflows. 
It is possible that the shielding gas is ionized and was therefore difficult to detect 
in the low signal-to-noise ratio {\sl Chandra} spectrum of \hs.
We investigated this  possibility  in section \S 2.1 by fitting the 5~ks {\sl Chandra} spectrum with a model consisting of a simple power law,  Galactic absorption of 3.94$\times$10$^{20}$~cm$^{-2}$ and intrinsic ionized absorption (model 2 of Table 2).
This model provides an acceptable fit in a statistical sense but does not provide a significant improvement  
(based on the $F$--test) over a model that includes a neutral absorber.
No useful constraints of the ionization parameter, $\xi$$_{\rm shield}$, and the hydrogen column density, N$_{\rm H,shield}$, are provided
in such a model (see panel b of Figure 3).

In Chartas et al. (2009a) we presented results from an X-ray survey of a sample of NAL quasars with high velocity UV outflows.
In contrast to what we find in BAL quasars we do not detect any significant excess intrinsic absorption in most NAL quasars.
Specifically, we found the intrinsic column densities of the X-ray absorbers in our sample of NAL quasars
were constrained to be less than a few $\times$ 10$^{22}$~cm$^{-2}$.
We also found that the distribution of  $\alpha_{ox}$ of the NAL quasars of our sample differs
significantly from those of BAL quasars (see Figures 2 and 3 of Chartas et al. 2009a), where,
$\alpha_{ox}$ is the optical-to-X-ray slope $\alpha_{ox}$ =$ 0.384\log(f_{2keV}/f_{\rm 2500\;\AA})$ (Tananbaum et al. 1979). 
 Since $\alpha_{ox}$ is known to correlate with UV luminosity
we also calculated the parameter $\Delta\alpha_{ox}$ = $\alpha_{ox}$ - $\alpha_{ox}(\ell_{\rm 2500\;\AA})$, where
$\alpha_{ox}(\ell_{\rm 2500\;\AA})$ is the expected $\alpha_{ox}$ for the
monochromatic luminosity at 2500{\rm \AA} (Strateva et al. 2005) based on an unbiased sample of
non-absorbed and radio-quiet AGN. 
$\Delta\alpha_{ox}$ is a proxy of X-ray weakness corrected for the dependence of $\alpha_{ox}$ on UV
luminosity.  We found that  the distribution of $\Delta\alpha_{ox}$ of the NAL quasars of our sample also differs
significantly from those of BAL quasars. 
The NAL quasars are not significantly absorbed in the X-ray band and the positive values of  $\Delta\alpha_{ox}$ suggest absorption in the UV band.
One plausible scenario that we presented to explain the  $\alpha_{ox}$ and $\Delta\alpha_{ox}$ distributions of the NAL quasars was that the
lines of sight towards the compact X-ray hot coronae of NAL quasars differ from the lines of sight towards their
UV emitting accretion disks. In Figure 10 we present a plausible
geometric configuration of a quasar outflow that can explain the non-detection of absorption from the shielding gas in the X-ray spectra of NAL quasars.  
These low values of N$_{\rm H}$ of NAL quasars are consistent with an outflow scenario in which
NAL quasars are viewed at smaller inclination angles than BAL quasars.
In this possible scenario, shielding gas may be present and intercepted along a
UV line of sight and thus aid in driving the UV wind but not
intercepted by the X-ray line of sight resulting in no significant absorption in the X-ray spectra of NAL quasars.

The confirmation of an absence of a mildly ionized absorber in the X-ray spectrum of \hs\ would place into question the 
hypothesis that a shielding gas is required to protect the X-ray wind from being over-ionized and efficiently drive the X-ray outflow.  
The non detection, however, of absorption in the X-ray spectrum of 
\hs\  would not rule out the presence of shielding gas along the UV line of sight to \hs.

In Figure 11 we show the VLT/UVES spectrum of \hs\  near the \CIV\ and \NV\ doublets, each form a different parcel of gas.
The observed wavelengths of the \CIV\ doublet ($\lambda\lambda$ 1548.20 $\rm \AA$, 1550.77 $\rm \AA$) implies that the 
absorber is outflowing with a speed of $v_{\rm C IV}$ $\sim$ 19,400 km~s$^{-1}$.  
The  \NV\ doublet  ($\lambda\lambda$ 1238.81~$\rm \AA$, 1242.80~$\rm \AA$) shows four outflowing components A, B, C, and D,
with velocities of $v_{\rm N V,A}$ = 789 km~s$^{-1}$, $v_{\rm N V,B}$ = 868 km~s$^{-1}$, 
$v_{\rm N V,C}$ = 1046 km~s$^{-1}$, and $v_{\rm N V,D}$ = 1198 km~s$^{-1}$, respectively. 

The \CIV\ and \NV\ NAL lines are likely 
intrinsic to the quasar outflow (Culliton et al., in prep.) based on 
a partial coverage analysis (e.g., Hamann et al. 1997; Misawa et al. 2007). 
Specifically, the partial covering analysis indicates that for at least one component 
in each CIV and NV system the covering fraction is more than 3 $\sigma$ below 1.
The rest-frame equivalent widths of the \CIV\ ($\lambda$ 1548.20 $\rm \AA$) and \NV\ ($\lambda$ 1238.81  $\rm \AA$)
lines are W$_{\rm C IV}$ = 0.11 $\rm  \AA$  and  W$_{\rm N V}$ = 0.3 $\rm \AA$ (sum of all kinematic components), respectively.

The outflow velocities of the UV and X-ray absorbing material differ significantly 
implying that the UV and X-ray absorption lines are produced by different absorbers. 
The ionization parameter of the gas that produces the \CIV\ and \NV\ absorption lines in the UV spectrum is considerably 
lower than that needed to explain the absorption lines in the X-ray spectrum. This reinforces the idea that
the X-ray and UV absorption lines originate from different absorbers.
The large velocities of the highly ionized X-ray absorber
are difficult to explain with radiation driving. Specifically, for large values of the ionization parameter of the absorber
fewer resonant transitions are available to absorb photons from the source, thus resulting in a weaker driving force.
Our recent analysis of a sample of mini-BAL quasars found that their X-ray absorption was weak or moderate 
with total neutral-equivalent column densities $N_{\rm H}$ $\simlt$ few $\times$ 10$^{22}$ cm$^{-2}$ (Hamann et al. 2013). 
This weak to moderate X-ray absorption might imply the lack of any significant shielding gas that could explain the 
observed moderate ionization levels (e.g., \CIV\ and \OVI) and the high velocities of the outflowing UV absorbing gas in these mini-BAL quasars.
However, as we mentioned earlier it is also possible that absorption from the shielding gas is 
undetected in the X-ray spectra of NAL quasars in a geometry in which the UV and X-ray lines of sight differ.
In this later case the shielding is likely not present along the X-ray line of sight but may be 
present along the UV line of sight.

\section{SUMMARY}
We have presented results from an analysis of a 5~ks  {\sl Chandra} observation of  
the lensed NAL AGN \hs\ ($z = 1.51$).
Despite the short exposure, the very large ($\sim$100) magnification factor offers
a unique opportunity for the study in X-rays of a borderline AGN/QSO (unlensed 2--10~keV 
luminosity of 5.8$\times$10$^{43}$ erg~sec$^{-1}$) near the 
cosmic peak of AGN activity ($z \sim 1-2$). The main conclusions of our spectral and spatial analyses are the following.\\

1. High-energy X-ray absorption lines  in the range of 1.5$-$5~keV (observed-frame)
are detected at the greater than 97\% confidence level in \hs.
Based on our fits that use the photoionization model XSTAR, the 
absorption lines centered near rest-frame energies of 7.7~keV  and 11.0~keV  (dominated by  \FeXXV\  ($1s^2-1s2p$))
are associated with absorbers outflowing at  0.13 $c$ and 0.41 $c$, respectively.
If we interpret the absorption line near rest-frame energy of 3.94~keV as arising from highly blueshifted \SXV\ ($1s^2-1s2p$)
(but at three times the solar abundance value) the implied outflow velocity of this absorber is also 0.41 $c$. \\

2.  Our estimated values of the mass-outflow rates for the outflowing components abs3 and abs4 are 
$\dot{M}_{\rm abs3}$ = 0.35$_{-0.25}^{+0.36}$ M$_{\odot}$~yr$^{-1}$ and $\dot{M}_{\rm abs4}$ = 1.1$_{-0.7}^{+0.9}$ M$_{\odot}$~yr$^{-1}$,
respectively. These mass-outflow rates are comparable to the estimated  accretion rate of \hs\  which we estimate
to be 1.8 $\times$ 10$^{-3}$($L_{44}$/$\eta$)$M_{\odot}$  yr$^{-1}$ $\sim$ 1 $M_{\odot}$ yr$^{-1}$,
where we assumed a typical accretion efficiency of $\eta$ = 0.1.
We conclude that the X-ray outflow of NAL AGN \hs\  distributes a significant amount of
accretion-disk material into the vicinity of the quasar central engine and into the host galaxy over the lifetime of its active phase.\\

3. Our estimated values of the fraction of the total bolometric energy released by \hs\ into the IGM in the form of kinetic
energy for components abs3 and abs4 are  
$\epsilon_{\rm k,abs3}$ = 0.04$_{-0.03}^{+0.07}$(0.1/$\lambda_{\rm Edd}$) and
$\epsilon_{\rm k,abs4}$ = 1.1$_{-0.6}^{+0.8}$(0.1/$\lambda_{\rm Edd}$), respectively.
An efficiency greater than the covering factor $f_{c}$ = 0.3--0.5 implies that radiation driving alone cannot explain the 
acceleration of this highly-ionized absorber. It is likely that magnetic driving is a significant 
contributor to the acceleration of this X-ray wind. \\

4. We detected a galaxy group $\sim$ 58\arcsec\ west of the center of \hs. By modeling the brightness profile of the galaxy group
we inferred the external shear at the location of \hs.
Our gravitational lens model of \hs\ consisted of a singular isothermal ellipsoid plus an external shear 
from the nearby galaxy group. We find that the magnifications of the images are $M_{\rm A}$=40, $M_{\rm B}$=38, $M_{\rm C}$=10, and $M_{\rm D}$=6, consistent with Assef et al. (2011). \\

5. Our analysis of the X-ray spectrum of \hs\ indicated the absence of any significant neutral absorber that could act to shield the outflow
of X-ray absorbing material.  The presence of outflows of X-ray absorbing gas at $v_{\rm X-ray}$ $\sim$ 0.41 $c$ 
suggest that a shielding gas is not required for the generation of the relativistic X-ray absorbing winds in \hs.
It is possible that the shielding gas is ionized and was therefore difficult to detect in the 5 ks {\sl Chandra} spectrum.
Deeper observations will be required to unambiguously reveal whether a neutral or moderately ionized shielding gas is present or not
along the X-ray line of sight.  If the UV and X-ray lines of sight are different, as in a plausible
outflow geometry (see Figure 9), it is possible that shielding is present along the UV line of sight but not
present along the X-ray line of sight.\\

6. Our analysis of the VLT/UVES spectrum of \hs\ shows highly blueshifted \CIV\ and \NV\ doublets implying 
outflow velocities of  $v_{\rm CIV}$ $\sim$ 19,400 km~s$^{-1}$ and up to $v_{\rm NV}$ $\sim$ 1,046 km~s$^{-1}$, respectively.
The significantly different UV and X-ray outflow velocities indicate that the UV and X-ray absorption lines are produced by different absorbers. \\

7. The detection of ultrafast winds in NAL quasars HS 0810+2554 (this study) and HS 1700+6414 (Lanzuisi et al., 2012), in addition to those 
detected in mini-BAL and BAL quasars, would suggest that the X-ray absorbing winds of quasars may have  
opening angles larger than previously thought, as large as $\sim$50\% if ubiquitous in all NAL and BAL quasars.
If confirmed, these winds would thus increase their implied feedback contribution to the surrounding ISM and IGM.

\acknowledgments
We acknowledge financial support from NASA via the Smithsonian Institution grants SAO GO1-12146B.
ME and JCC acknowledge financial support from NSF grant AST-0807993.

\clearpage
\begin{table}
\caption{Log of Observation of Quasar \hs\ } 
\scriptsize
\begin{center}
\begin{tabular}{cccccccccc}
                                   &                           &   & & & &   \\ \hline\hline
                                   &  {\it Chandra}  & Exposure            &                                                               &                                                            &                                                             &  \\
 Observation           &   Observation  &   Time                  &  $N_{\rm A}$$\tablenotemark{a}$  & $N_{\rm B}$$\tablenotemark{a}$ & $N_{\rm C}$$\tablenotemark{a}$ & $N_{\rm D}$ $\tablenotemark{a}$   \\
     Date                    &         ID             &   (s)                     & counts                                                 & counts                                               & counts                                                & counts   \\
&   &    &   & & &\\
\hline
2002 January  30  &  3023               &  4894                  &  290$^{+33}_{-27}$                             & 239$^{+41}_{-43}$                           & 145$^{+15}_{-18}$            & 34$^{+9}_{-8}$   \\

\hline \hline
\end{tabular}
\end{center}
${}^{a}${Source counts for events with energies in the 0.2--10~keV band.}\\
\end{table}

\clearpage
\begin{table}
\caption{Results from Fits  to the \chandra\ Spectrum of \hs}
\scriptsize
\begin{center}
\begin{tabular}{cccc}
 & & & \\ \hline\hline
\multicolumn{1}{c} {Model$^{a}$} &
\multicolumn{1}{c} {Parameter$^{b}$} &
\multicolumn{1}{c} {Fitted Values$^{c}$} & 
\multicolumn{1}{c} {Fitted Values$^{d}$}  \\
    &              &     &                          \\
  ${1}$ &$\Gamma$     &  1.72$_{-0.12}^{+0.12}$  &    1.65$_{-0.09}^{+0.12}$ \\
    &   $N_{\rm H}$ & $<$ 1.2$\times$10$^{21}$~cm$^{-2}$   &  $<$ 6$\times$10$^{20}$~cm$^{-2}$\\
 &$\chi^2/{\nu}$  & 65.9/81 &  245.8/296\\
&$P(\chi^2/{\nu})$$^{e}$ & 8.9~$\times$~10$^{-1}$   & --  \\
  &                                                             &               &   \\
    &                                         &                                    &       \\
${2}$  &$\Gamma$ & 1.74$_{-0.12}^{+0.12}$           &    1.72$_{-0.12}^{+0.08}$      \\
  &   $N_{\rm H}$ &   ${}^{f}$  &  ${}^{f}$  \\
  &    $log\xi_{}$            & ${}^{f}$  & ${}^{f}$  \\
  &   $\chi^2/{\nu}$   & 65.8/80                  &   243.8/295         \\
  &  $P(\chi^2/{\nu})$$^{e}$ & 8.7~$\times$~10$^{-1}$           &  --    \\ 
    &  &                     &           \\
${3}$ &  $\Gamma$          & 1.50$_{-0.12}^{+0.14}$                  &    1.57$_{-0.12}^{+0.13}$     \\
&   $N_{\rm H}$                 & $<$ 0.93$\times$10$^{21}$~cm$^{-2}$  &  $<$ 0.68 $\times$10$^{21}$~cm$^{-2}$\\
  &  E$_{\rm abs1}$          & 3.94$_{-0.31}^{+0.17}$~keV                      &    3.94$_{-0.58}^{+0.15}$~keV  \\
  &  $\sigma_{\rm abs1}$ & 0.15$_{-0.15}^{+2.02}$~keV                      &    0.17$_{-0.17}^{+0.84}$~keV \\
  &  EW$_{\rm abs1}$       & $-$210$_{-110}^{+102}$~eV                     & $-$243$_{-91}^{+97}$~eV \\
  &  E$_{\rm abs2}$           & 4.96$_{-0.09}^{+0.09}$~keV                      &  5.01$_{-0.14}^{+0.09}$~keV \\
  &  $\sigma_{\rm abs2}$ & $<$ 0.27~keV                                                 &  $<$ 0.26~keV  \\
  &  EW$_{\rm abs2}$       &$-$246$_{-109}^{+117}$~eV                       &    $-$235$_{-85}^{+89}$~eV             \\
   &  E$_{\rm abs3}$         & 7.74$_{-0.24}^{+0.21}$~keV                        &    7.76$_{-0.37}^{+0.35}$~keV        \\
  &  $\sigma_{\rm abs3}$ & $ < $ 0.6~keV                                                 &  0.30$_{-0.17}^{+0.45}$~keV     \\
  &  EW$_{\rm abs3}$       & $-$495$_{-171}^{+287}$~eV                      &    $-$413$_{-281}^{+223}$~eV         \\
  &  E$_{\rm abs4}$           & 11.00$_{-0.77}^{+0.82}$~keV                    &     11.17$_{-0.85}^{+0.90}$~keV     \\
  &  $\sigma_{\rm abs4}$  & 0.95$_{-0.94}^{+1.16}$~keV                      & 0.97$_{-0.77}^{+0.90}$~keV   \\
  &  EW$_{\rm abs4}$        & $-$1400$_{-750}^{+910}$~eV                   &     $-$1257$_{-894}^{+770}$~eV        \\
  & $\chi^2/{\nu}$                & 45.6/69                                                            & 222.8/284  \\
 &$P(\chi^2/{\nu})$$^{e}$ & 9.9~$\times$~10$^{-1}$                              &  --                          \\
  &           &                                    & \\
${4}$  &$\Gamma$            & 1.75$_{-0.2}^{+0.1}$                                                         &  1.76$_{-0.1}^{+0.1}$           \\
  &   $N_{\rm Habs3}$        & 1.7$_{-0.6}^{+2.1}$ $\times$ 10$^{23}$~cm$^{-2}$  &   1.0$_{-0.8}^{+1.0}$ $\times$ 10$^{23}$~cm$^{-2}$  \\
  &    $log\xi_{abs3}$          & 3.37$_{-0.2}^{+0.2}$~erg~cm~s$^{-1}$                       &  3.2$_{-0.1}^{+0.5}$~erg~cm~s$^{-1}$ \\
   &   $z_{abs3}$                  &  1.21$_{-0.16}^{+0.22}$                                                   &  1.20$_{-0.3}^{+0.16}$ \\
    &   $N_{\rm Habs4}$      &  1.62$_{-0.9}^{+1.6}$ $\times$ 10$^{23}$~cm$^{-2}$  &  1.64$_{-0.6}^{+1.9}$ $\times$ 10$^{23}$~cm$^{-2}$  \\
  &    $log\xi_{abs4}$          & 3.06$_{-0.2}^{+0.2}$~erg~cm~s$^{-1}$                        &  3.06$_{-0.02}^{+0.07}$~erg~cm~s$^{-1}$  \\
   &   $z_{abs4}$                  &  0.63$_{-0.09}^{+0.08}$                                                   &  0.62$_{-0.03}^{+0.06}$ \\
  &   $\chi^2/{\nu}$               & 60.4/76                                                                                 &  235/291                    \\
&  $P(\chi^2/{\nu})$$^{e}$ & 9.1~$\times$~10$^{-1}$                                                   &  --      \\ 
    &  &                      &          \\
  \hline \hline
\end{tabular}
\end{center}
\noindent
${}^{a}$ Model 1 consists of a power law and neutral absorption at the source. 
Model 2 consists of a power law and ionized absorption at the source.
Model 3  consists of a power law, neutral absorption at the source, and Gaussian absorption lines at the source.
Model 4  consists of a power law, neutral absorption at the source and two outflowing ionized absorbers at the source.
All model fits include the Galactic absorption toward the source (Dickey \& Lockman 1990).\\
${}^{b}$All absorption-line parameters are calculated for the rest frame.\\
${}^{c}$Spectral fits were performed using the $\chi^{2}$ statistic and all errors are for 90\% confidence unless mentioned otherwise with all
parameters taken to be of interest except absolute normalization.\\
${}^{d}$Spectral fits were performed using the Cash statistic and all errors are for 90\% confidence unless mentioned otherwise with all
parameters taken to be of interest except absolute normalization.\\
${}^{e}$$P(\chi^2/{\nu})$ is the probability of exceeding $\chi^{2}$ for ${\nu}$ degrees of freedom
if the model is correct.\\
${}^{f}$ No useful constraints are obtained for this parameter.
\end{table}

\clearpage 
\begin{table}
\caption{Hydrogen Column Densities, Outflow Velocities,  Mass-Outflow Rates and Efficiencies of Outflow in \hs}
\scriptsize
\begin{center}
\begin{tabular}{ccccc}
 & & &  & \\ \hline\hline
 Component & N$_{\rm H}$ & $v_{\rm abs}$  &  $\dot{M}$ & $\epsilon_{\rm k}$  \\
                 &           cm$^{-2}$  & $c$ & (M$_{\odot}~yr^{-1}$)  &       \\
&    &    &    &  \\
\hline
&    &   &  &  \\
$abs3$         & $1.7_{-0.6}^{+2.1}$ $\times$ 10$^{23}$ &$0.13_{-0.09}^{+0.07}$& $0.35_{-0.25}^{+0.36} $        & $0.04_{-0.03}^{+0.07}$ \\
$abs4$          & $1.6_{-0.9}^{+1.6}$ $\times$ 10$^{23}$ &$0.41_{-0.04}^{+0.05}$& $1.1_{-0.7}^{+0.9}$         & $1.0_{-0.6}^{+0.8}$ \\
\hline
&    &    & & \\
\hline \hline
\end{tabular}
\end{center}
\tablecomments{ The parameters assumed in the estimates of the kinematics of the absorbers are taken
from the best-fit values obtained from model 4 of Table 2. 
The bolometric luminosity of \hs\ is $L_{\rm bol}$=2.4~$\times$~10$^{45}$(94/$\mu$)($\kappa_{2-10keV}$/30)~erg~s$^{-1}$, 
where $\mu$ is the magnification factor and $\kappa_{2-10keV}$ is the bolometric correction factor. 
The efficiency of the outflow is defined as $\epsilon_{\rm k} = (1/2)\dot{M}v^{2}/L_{\rm bol}$.\\
 \\
}
\end{table}




\clearpage

  \begin{figure}
   \includegraphics[width=16cm]{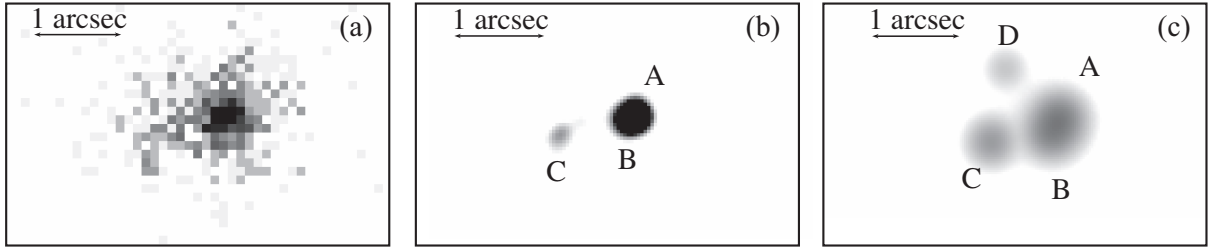}
        \centering
\caption[]{
Images of the 2002 January 30 {\sl Chandra} observation of quasar \hs. (a) Raw 0.2-10~keV image binned with a bin size of 0\sarc1 on a side. (b) Deconvolved image (0.2$-$10~keV band) of \hs. Images A and B are not resolved and image D is not reconstructed likely due to the low S/N of the {\sl Chandra} observation. (c) Best fit PSF model to the same observation of \hs.
In all panels north is up and east is to the left.}
\label{fig:images}
\end{figure}

\clearpage
 \begin{figure}
   \includegraphics[width=16cm]{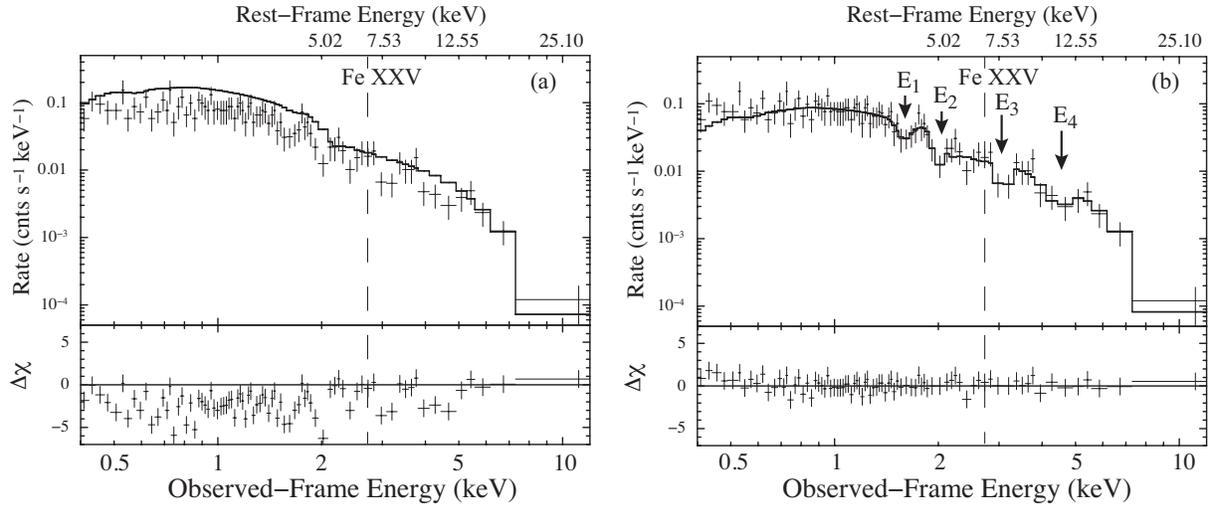}
        \centering
\caption[]{(a) The top panel shows the \chandra\ spectrum of
\hs\ fit with Galactic absorption and a 
power-law model to events with energies lying within 
the observed-frame range 5--10~keV. 
The lower panel shows the residuals of the fit in units of 1$\sigma$ deviations.
Several absorption features within the observed-frame range of 1.5--5.0~keV are noticeable in the residuals plot. 
The top panel (b) shows the same data shown in panel (a)  overplotted with the best-fit model
taken from \hbox{model 3} of \hbox{Table 2}. 
The arrows indicate the best-fit energies of the absorption lines of the four components obtained in the fit that used model 3 of Table 2.
The lower panels of (b) shows the reduced residuals 
of these fits. }
\label{fig:images}
\end{figure}

\clearpage
 \begin{figure}
   \includegraphics[width=16cm]{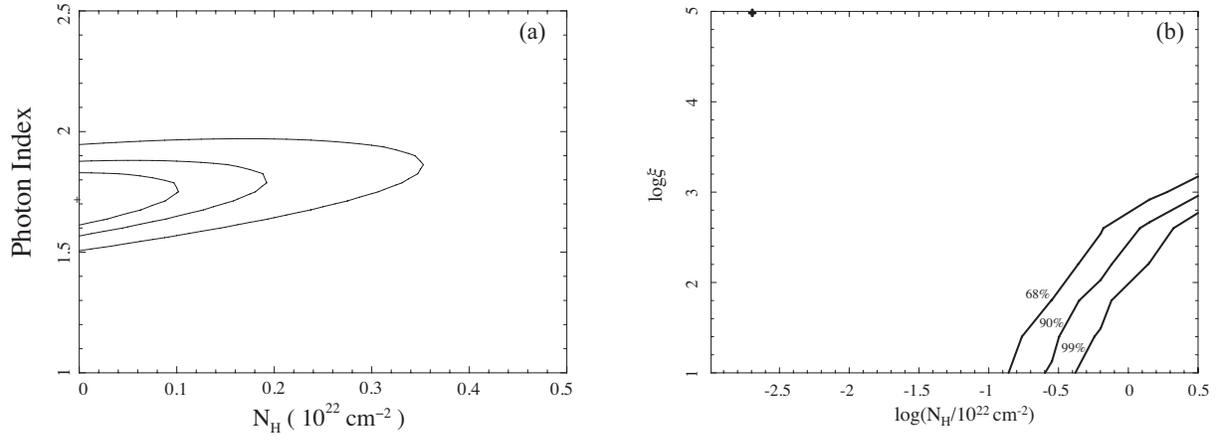}
        \centering
\caption[]{68\%, 90\%, and 99\%  $\chi^2$ confidence contours of (a)
 N$_{\rm H}$ versus photon index obtained in the fit that used model 1 of Table 2,  and, (b) 
N$_{\rm H}$ versus the ionization parameter obtained in the fit that used model 2 of Table 2. }
\label{fig:images}
\end{figure}

\clearpage
 \begin{figure}
   \includegraphics[width=16cm]{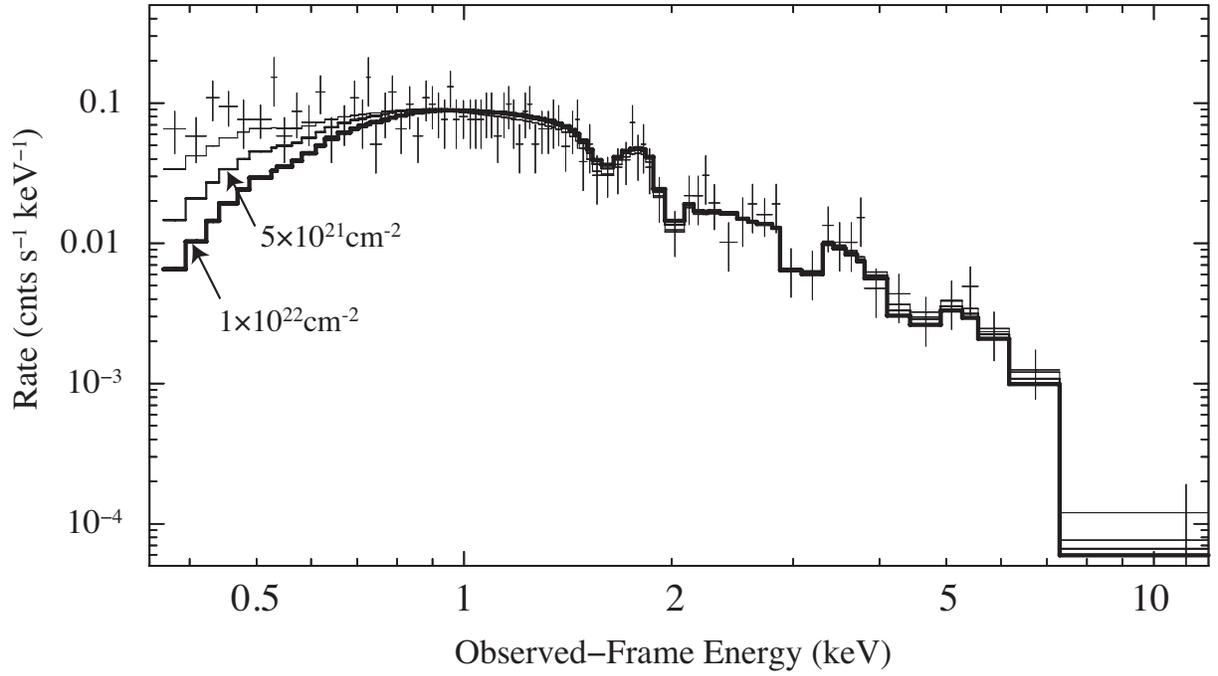}
        \centering
\caption[]{ 
The best-fit models (based on  model 3 of Table 2) where we forced the intrinsic column density to have a fixed value
of $N_{\rm H} = 5 \times 10^{21}~cm^{-2}$(case 1) and $N_{\rm H} = 1 \times 10^{22}~cm^{-2}$(case 2), 
over-layed with the best-fit model 3 of Table 2, where $N_{\rm H}$ was allowed to be a free parameter.
Hydrogen column densities assumed for cases 1 and 2 would lead to poor fits with significant
residuals in the 0.35--0.8 keV observed-frame energy band.
}
\label{fig:images}
\end{figure}

\clearpage
 \begin{figure}
   \includegraphics[width=16cm]{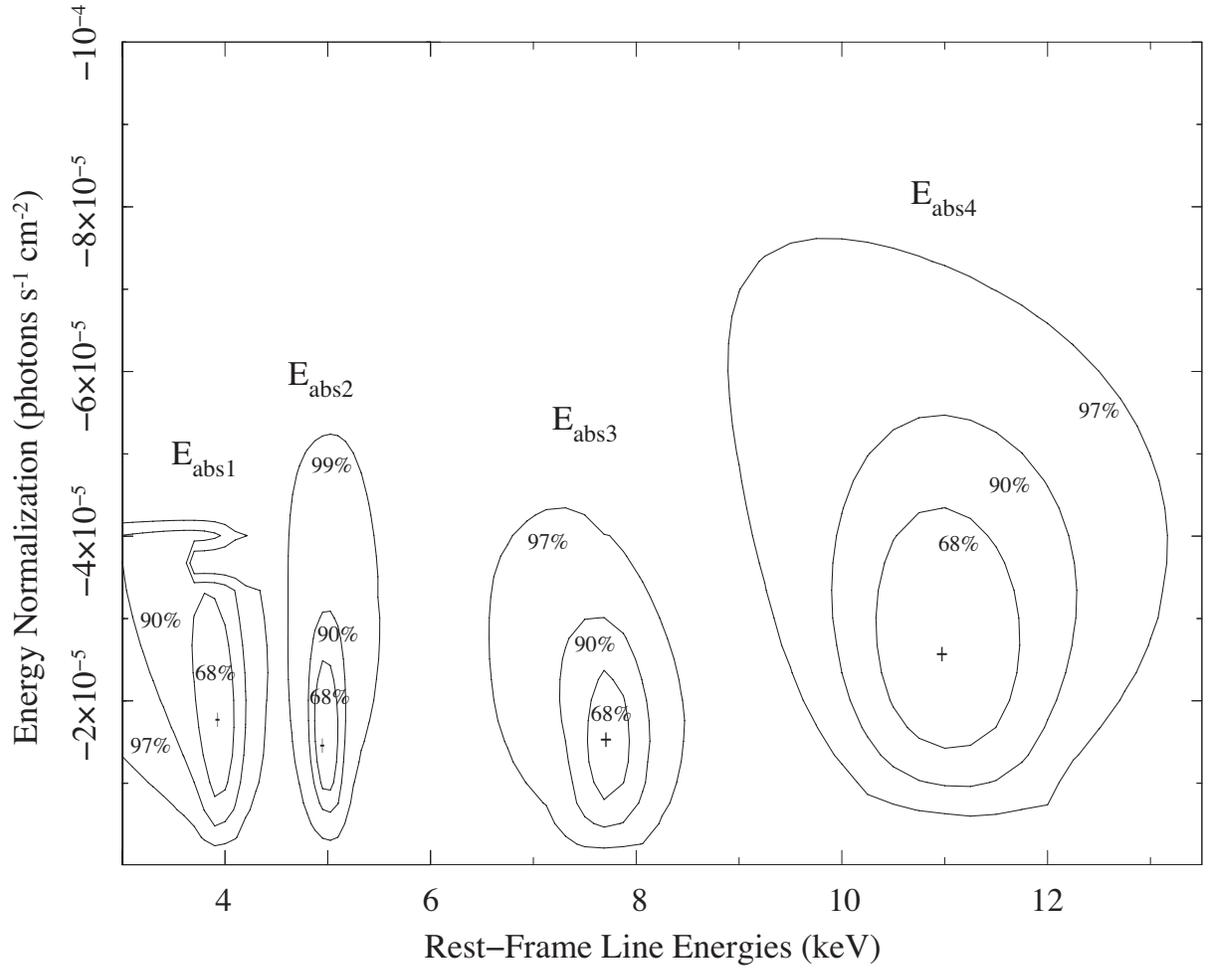}
        \centering
\caption[]{$\chi^2$ confidence contours 
between the normalizations of the absorption lines at  $E_{\rm abs1}$, $E_{\rm abs2}$, $E_{\rm abs3}$, and $E_{\rm abs4}$ and their respective energies.
The 99\% confidence contours, of $E_{\rm abs1}$, $E_{\rm abs3}$, and $E_{\rm abs4}$ are erratic and not closed at the
99\% level and are therefore not displayed.}
\label{fig:images}
\end{figure}

\clearpage
 \begin{figure}
   \includegraphics[width=16cm]{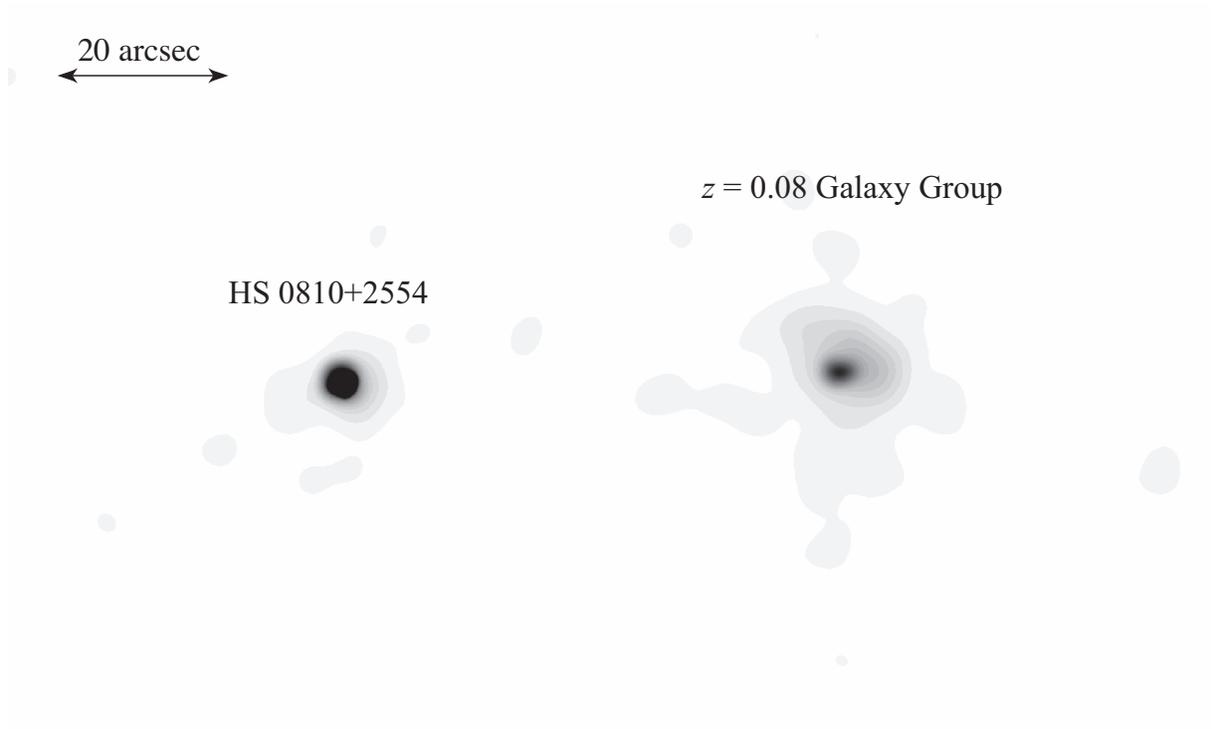}
        \centering
\caption[]{Adaptively smoothed image of the \chandra\ observation of \hs. Extended X-ray emission is detected 
58 \arcsec\ west of \hs\ at 8 13 27, $+$25 45 4.5 (J2000). Our analysis indicates that the origin of the extended emission is 
likely a galaxy group at $z \sim 0.08$.}
\label{fig:images}
\end{figure}

\clearpage
 \begin{figure}
   \includegraphics[width=16cm]{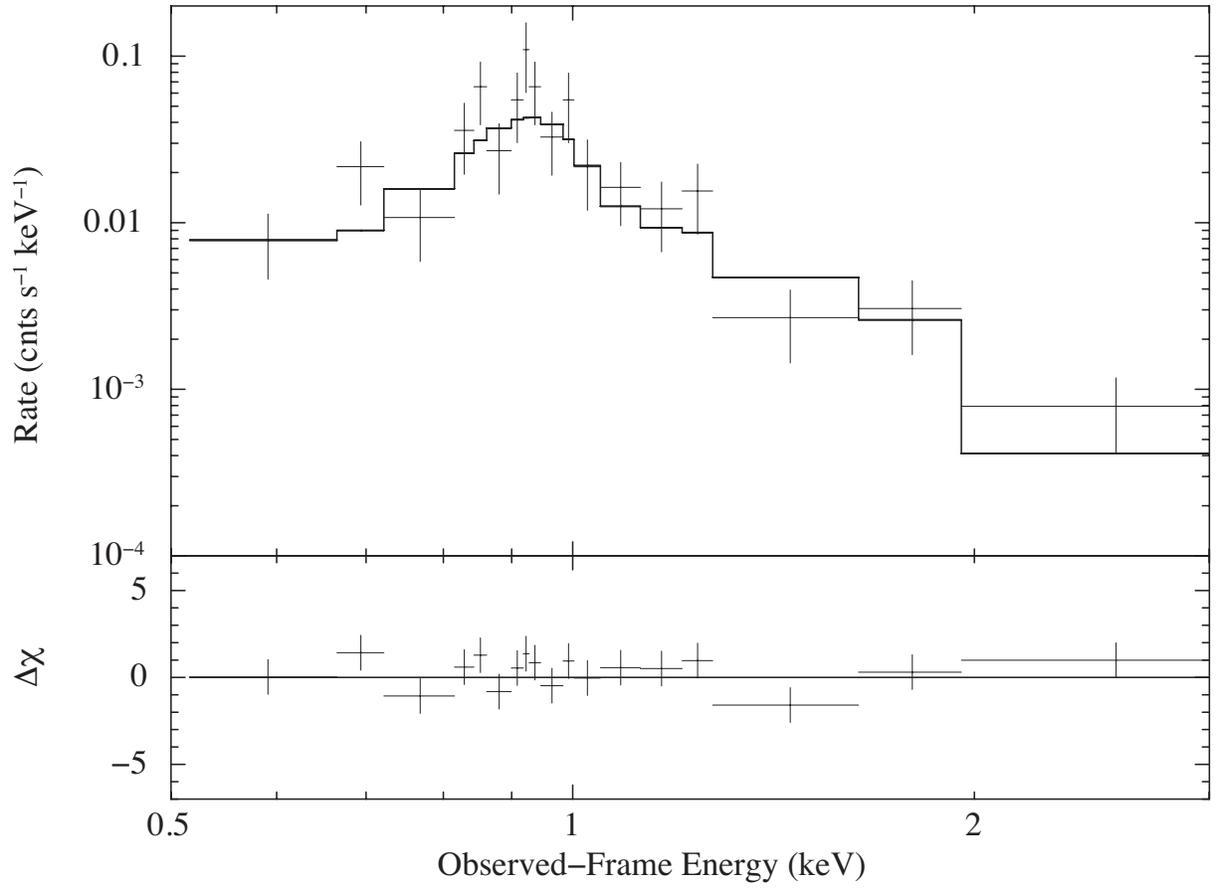}
        \centering
\caption[]{The X-ray spectrum of the galaxy group centered 58\arcsec\ west of  \hs\ along with the 
best-fit thermal {\it mekal} model. The model is consistent
with a plasma temperature of $\sim$ 1.0 keV.  (Lower panel) Residuals in units of standard deviations with error bars of size 1$\sigma$. }
\label{fig:images}
\end{figure}

\clearpage
 \begin{figure}
   \includegraphics[width=16cm]{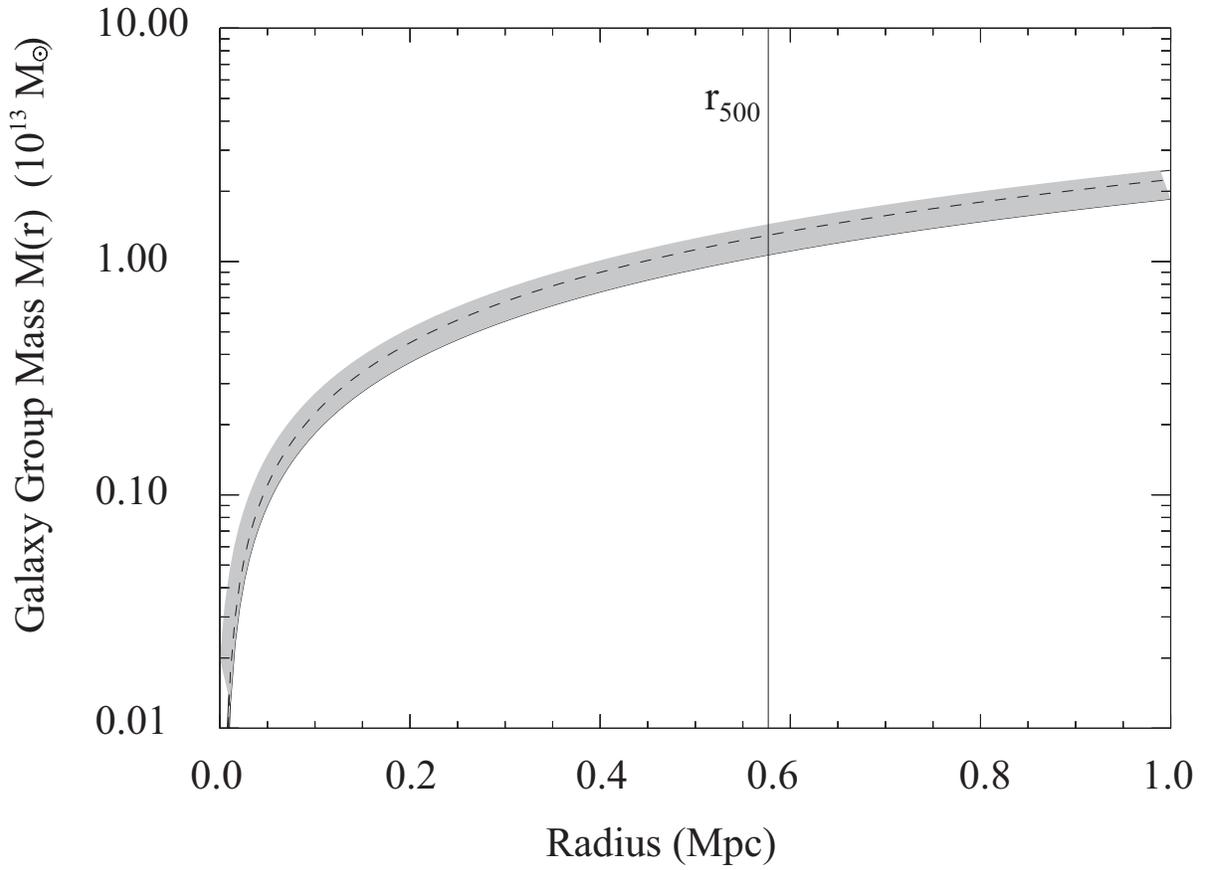}
        \centering
\caption[]{The total mass of the galaxy group centered 58\arcsec\ west of  \hs\  within
a radius $r$. The dashed line corresponds to the best-fit spatial
and spectral parameters. The shaded region indicates the allowable 
range of the galaxy group mass for the estimated uncertainties
of the best-fit spatial and spectral parameters. The vertical solid line corresponds to the 
radius where the mean density of the group is 500 times the critical density of the Universe. }
\label{fig:images}
\end{figure}

\clearpage
 \begin{figure}
   \includegraphics[width=16cm]{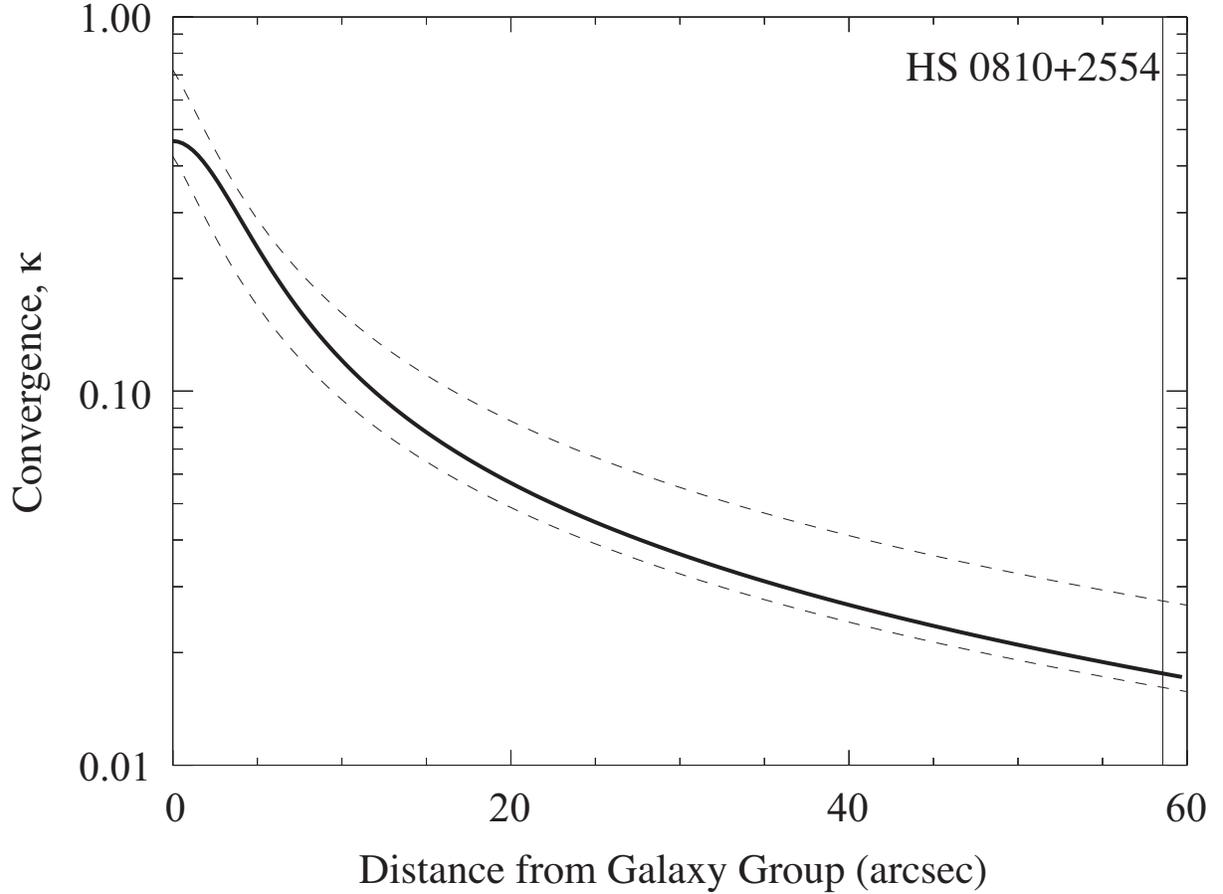}
        \centering
\caption[]{The convergence parameter $\kappa$ of the
galaxy group as a function of distance from the group center.
The thick solid line corresponds to the best-fit spatial and spectral parameters.
The largest contributor to the uncertainty in the present 
measurement of $\kappa(x)$ is the weak constraint on the temperature of the group.
To illustrate this weakness we have plotted the uncertainty in
$\kappa(x)$ assuming 90\% (dashed lines) 
confidence intervals for the temperature. 
We also chose cluster limits ranging from 0.7$r_{500}$ and 1.4$r_{500}$,
where $r_{500}$ is the radius in which the mean over-density is 500,
and $r_{500}$ = 1.69 $h^{-1}_{70}$ Mpc (T$_{X}$/10~keV)$^{1/2}$ $\sim$ 0.55 $h^{-1}_{70}$ Mpc
(Mohr, Mathiesen, \& Evrard, 1999). The solid vertical line indicates the 
distance of the center of \hs\ from the galaxy group center.}
\label{fig:images}
\end{figure}

\clearpage
 \begin{figure}
   \includegraphics[width=16cm]{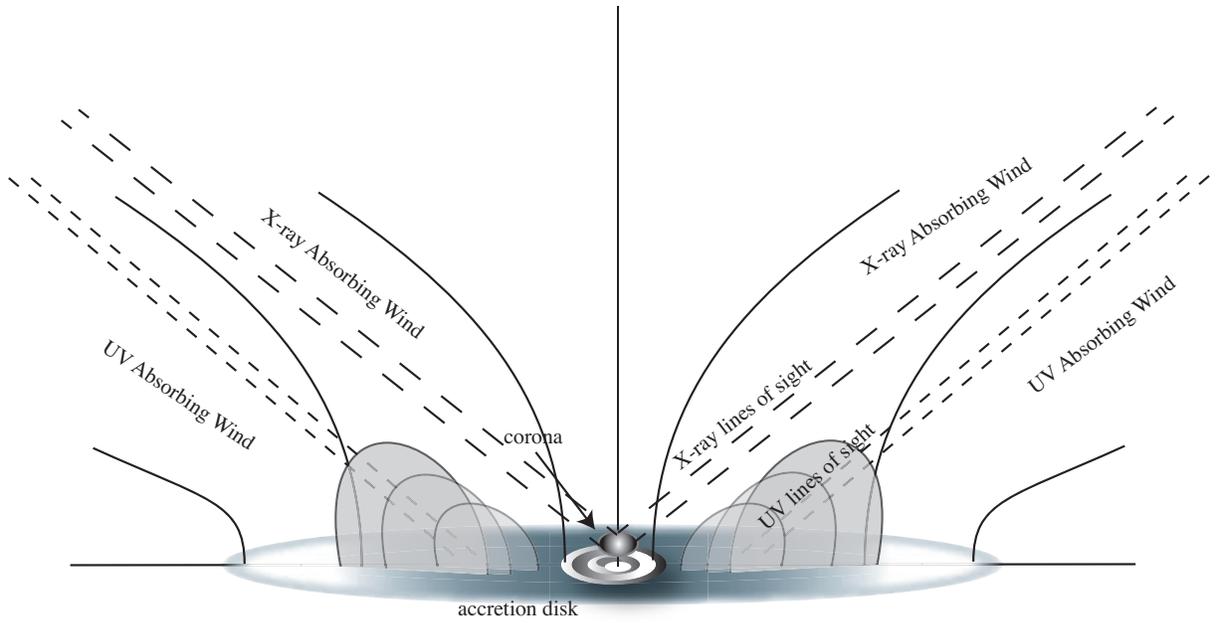}
        \centering
\caption[]{
Schematic diagram of a plausible geometry for the accretion disk and associated outflow in quasars. 
X-ray lines of sight originating from the corona are indicated with solid lines and UV lines of sight originating from the accretion disk are indicated with dashed lines.The failed wind that may act as a shielding gas is illustrated with streamlines colored in grey. For small inclination angles 
the X-ray lines of sight do not intercept the shielding gas whereas the UV lines of sight do. }
\label{fig:images}
\end{figure}

\clearpage
 \begin{figure}
   \includegraphics[width=11cm]{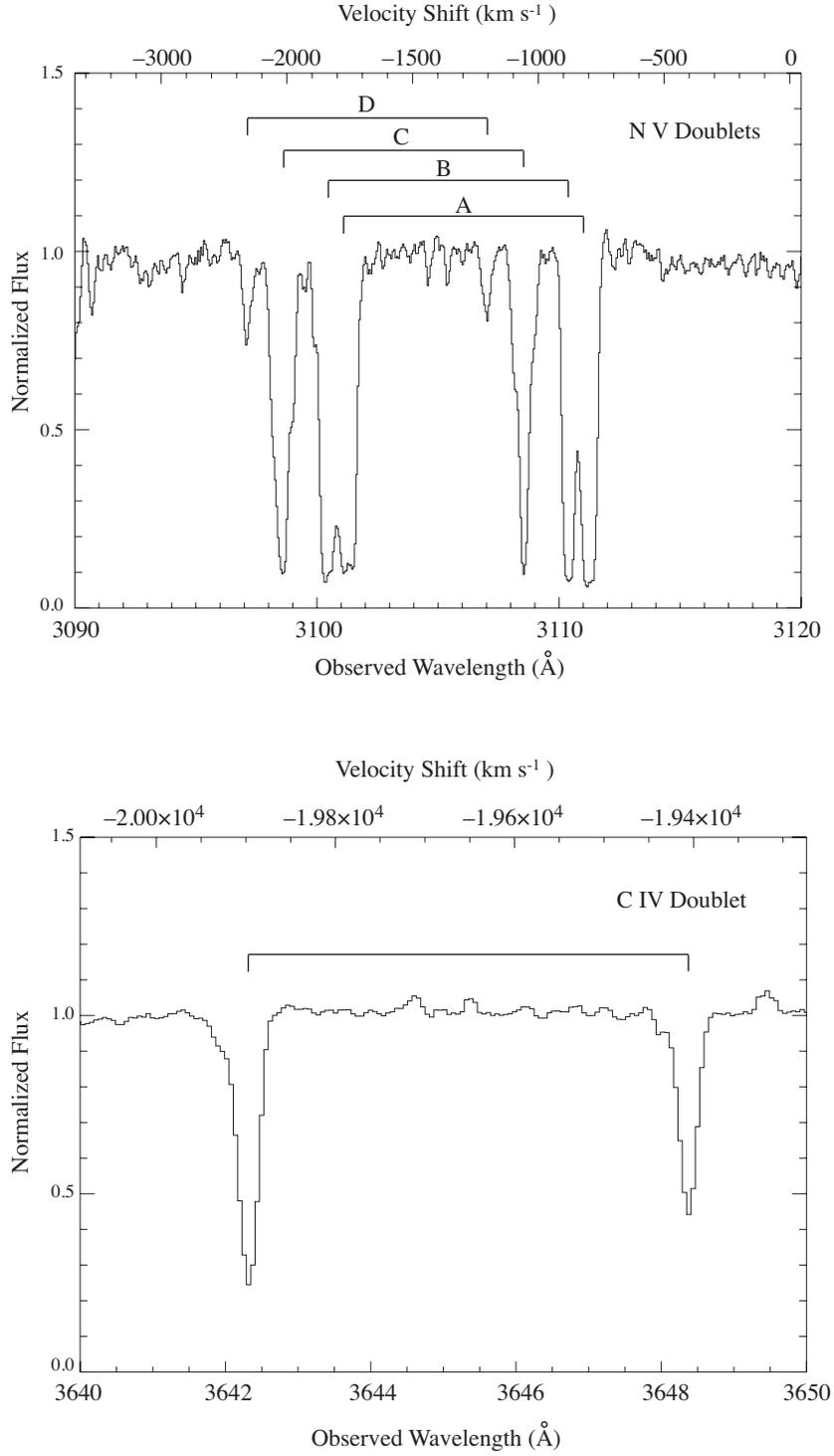}
        \centering
\caption[]{
The UV spectrum of \hs\ taken with the VLT/UVES showing blueshifted,   
(a) \NV\ doublets with four components labelled A, B, C, D outflowing at  
$v_{\rm NV ,A}$ = 789~km~~s$^{-1}$,  $v_{\rm NV,B}$ = 868~km~~s$^{-1}$, $v_{\rm NV,C}$ = 1046~km~~s$^{-1}$, and
$v_{\rm NV,D}$ = 1198 km~s$^{-1}$, respectively, and,
(b) \CIV\ doublet outflowing at $v_{\rm CIV}$ = 19,400~km~~s$^{-1}$.}
\label{fig:images}
\end{figure}

\end{document}